\documentclass[aps,pre,amsmath,amsfonts,amssymb,superscriptaddress,preprint]{revtex4-1}

\usepackage{amsmath}
\usepackage{amsfonts}
\usepackage{enumitem}
\usepackage{yfonts}
\usepackage{mathrsfs}

\usepackage{amssymb}
\usepackage[english]{babel}
\usepackage[pdftex]{graphicx}
\usepackage{tikz}
\usepackage{rotating}
\usepackage[pdftex,hypertexnames=true,linktocpage=true,breaklinks=true]{hyperref} 
\usepackage[hyphenbreaks]{breakurl}
\hypersetup{colorlinks=true,linkcolor=blue,anchorcolor=blue,citecolor=magenta,filecolor=blue,urlcolor=blue,bookmarksnumbered=false,pdfview=FitB}

\usepackage{esvect}

\newcommand{\bea}{\begin{eqnarray}}
\newcommand{\eea}{\end{eqnarray}}
\newcommand{\bml}{\begin{mathletters}}
\newcommand{\eml}{\end{mathletters}}

\newcommand{\f}{\begin{equation}}
\newcommand{\ff}{\end{equation}}

\newcommand{\ba}{\begin{array}}
\newcommand{\ea}{\end{array}}
\newcommand{\bena}{\begin{eqnarray}}
\newcommand{\eena}{\end{eqnarray}}
\newcommand{\bdis}{\begin{displaymath}}
\newcommand{\edis}{\end{displaymath}}
\newcommand{\bit}{\begin{itemize}}
\newcommand{\eit}{\end{itemize}}
\newcommand{\ben}{\begin{enumerate}}
\newcommand{\een}{\end{enumerate}}

\newcommand{\Si}{\Sigma}

\newcommand{\ngV}{\| \nabla V \|}

\newcommand{\de}{\partial}

\everymath{\displaystyle}
\newcommand{\Hess}{\mathrm{Hess}}
\newcommand{\bi}{\begin{itemize}}
\newcommand{\ei}{\end{itemize}}

\newcommand{\intSigmaV}[2]{\int_{\Sigma^{#1}_{#2}}}

\newcommand{\affF}{DSFTA, University of Siena, Via Roma 56, 53100 Siena, Italy}
\newcommand{\affE}{INFN Sezione di Perugia, I-06123 Perugia, Italy}

\newcommand{\affA}{Aix-Marseille University, Marseille, France}
\newcommand{\affB}{CNRS Centre de Physique Th\'eorique UMR7332,
13288 Marseille, France}
\newcommand{\affC}{QSTAR \& CNR - Istituto Nazionale di Ottica, largo E. Fermi 6, 50125 Firenze, Italy}
\newcommand{\affD}{Dipartimento di Fisica Universit\`a di Firenze, and
I.N.F.N., Sezione di Firenze, via G. Sansone 1, I-50019 Sesto Fiorentino, Italy }
\newcommand{\affN}{Department of Physics and Sciences of Materials, University of Luxembourg, Luxembourg}
\newcommand{\affM}{Quantum Biology Lab, Howard University, 2400 6th St NW, Washington, DC 20059, USA }
\begin{document}

\newtheorem{theorem}{Theorem}
\newtheorem{lemma}{Lemma}
\newtheorem{remark}{Remark}
\newcommand{\non}{\nonumber}
\newcommand{\be}{\begin{equation}}
\newcommand{\ee}{\end{equation}}
\newcommand{\bq}{\begin{eqnarray}}
\newcommand{\eq}{\end{eqnarray}}
\newcommand{\lps}{\langle}
\newcommand{\rps}{\rangle}
\newcommand{\vb}{\bar{v}}
\newcommand{\bv}{\bar{v}}
\newcommand{\bfit}{\textit{\textbf  } }
\newcommand{\D}{\mathrm{d}}
\newcommand{\LevelSetfunc}[2]{\Sigma_{#1}^{#2}}
\newcommand{\Mfunc}[2]{M_{#1}^{#2}}
\everymath{\displaystyle}

\title{Topological Theory of Phase Transitions}

\date{\today}
\author{Matteo Gori}
\email{gori6matteo@gmail.com}
\affiliation{\affM}\affiliation{\affN}

\author{Roberto Franzosi}
\email{roberto.franzosi@ino.it}
\affiliation{\affF}\affiliation{\affE}\affiliation{\affC}

\author{Giulio Pettini}
\email{pettini@fi.unifi.it}
\affiliation{ \affD}

\author{Marco Pettini}
\email{pettini@cpt.univ-mrs.fr}
\affiliation{\affA}\affiliation{\affB}

\begin{abstract}
The investigation of the Hamiltonian dynamical counterpart of phase transitions, combined with the Riemannian geometrization of Hamiltonian dynamics, has led to a preliminary formulation of a differential-topological theory of phase transitions. In fact, in correspondence of a phase transition there are peculiar geometrical changes of the mechanical manifolds that are found to stem from changes of their topology. These findings, together with two theorems, have  suggested that a topological theory of phase transitions can be formulated  to go beyond the limits of the existing theories. Among other  advantages, the new theory applies to phase transitions in small $N$ systems (that is, at nanoscopic and mesoscopic scales), and in the absence of symmetry-breaking. However, the  preliminary version of the theory was incomplete and still falsifiable by counterexamples. 
The present work provides a relevant leap forward leading to an accomplished development of the topological theory of phase transitions paving the way to further developments and applications of  the theory that can be no longer hampered. 
\end{abstract}
\maketitle
\section{Introduction}
Phase transitions phenomena are ubiquitous in nature at very different scales in space and in energy.  
Therefore, from the theoretical viewpoint, understanding their origin, and the way of classifying them, is of central interest.
In spite of a huge literature on this topic, a general theory is still lacking.
In the framework of Landau’s phenomenological theory, phase transitions are generally related to the mechanism of spontaneous symmetry breaking.
However, Landau’s theory is not  all-encompassing. Indeed, many systems do not fit in this theory and undergo a phase transition lacking  spontaneous symmetry breaking and lacking an order parameter. Some notable  examples are :  Kosterlitz-Thouless transitions after Mermin-Wagner theorem, systems with local gauge symmetries after Elitzur’s theorem, liquid-gas transitions, transitions in supercooled glasses and liquids, transitions in amorphous and disordered systems, folding transitions in homopolymers and proteins. 
Furthermore, to account for the loss of analyticity of thermodynamic observables, the mathematical description of phase transitions requires the limit of an infinite number of particles (thermodynamic limit) as is the case of the Yang-Lee theory \cite{YL} and  of the Dobrushin-Lanford-Ruelle theory \cite{DLR}.  However, the contemporary research on nanoscopic and mesoscopic systems,  on the biophysics of polymers \cite{bachmann,PRL-Bachmann}, on Bose-Einstein condensation, Dicke superradiance in microlaser, superconducting transitions in small metallic objects, tackles transition phenomena in systems of finite - and often very small - number of particles.

Within all the hitherto developed theoretical frameworks, also including the monumental theory of Renormalization Group and critical phenomena, it is assumed that the primitive object at the grounds of a theory is a given statistical measure; schematically: the gran-canonical measure in the old Yang-Lee theory, the canonical measure in the Dobrushin-Lanford-Ruelle theory, and the microcanonical measure in a still somewhat open and more recent approach \cite{gross,bachmann,PRL-Bachmann}. However, there are several general results suggesting that the possibility for a system to undergo a phase transition depends on some measure-independent properties, as is its spatial dimension, the dimensionality of its order parameter, the range of its interactions, the symmetry group (discrete or continuous) of its Hamiltonian. This hints at the possibility that the same information might be encoded already at a more fundamental level completely determined by the internal interactions of a system, interactions described by their potential function. 

Therefore, looking for generalisations of the existing theories is a well motivated and timely purpose. The present paper puts forward a new starting point  for a line of thought initiated several years ago and based on a variety of results which hitherto did not appear to fit in a coherent theoretical framework. The central idea of this line of thought is that the singular energy dependence of the thermodynamic observables at a phase transition is the ”shadow” of some adequate \textit{change of topology} of the energy level sets in phase space (or of the potential level sets in configuration space, as well). 

\subsection{Why topology}
\smallskip

 
Recently, the study of equilibrium phase transitions in the microcanonical ensemble has attracted increasing interest, being very important in presence of ensemble inequivalence, when only the microcanonical ensemble gives the correct results.  Two complementary approaches have been undertaken. One of these is of a statistical kind \cite{gross,bachmann}, recently summarized in a very interesting, powerful and rich classification of microcanonical phase transitions by M.Bachmann in Ref.\cite{PRL-Bachmann}. On another side, as the ergodic invariant measure of nonintegrable Hamiltonian systems is the microcanonical measure, the other approach resorts to the study of Hamiltonian dynamics of systems undergoing phase transitions. 
This dynamical approach brings about  interesting novelties with respect to the standard studies of phase transitions, eventually leading to the Topological Hypothesis (TH) through the following logical chain.
The dynamics of a generic system of  $N$ degrees of freedom described by a Hamiltonian $H = \frac{1}{2}\sum_{i=1}^N p_i^2 + V(q_1,\ldots,q_N)~,$
or equivalently  by the corresponding Lagrangian function $L = \frac{1}{2}\sum_{i=1}^N {\dot q}_i^2 - V(q_1,\ldots,q_N)~,$ is chaotic. The degree of chaoticity of the dynamics is measured by the largest Lyapunov exponent, a new observable that has been proven useful to characterize phase transitions from a dynamical viewpoint \cite{book}. Then, the explanation of the origin of Hamiltonian chaos - encompassing the computation of the largest Lyapunov exponent -  proceeds by identifying a Hamiltonian flow with a geodesic flow of an appropriate Riemannian differentiable manifold. This differential geometric framework is given by configuration space endowed with the non-Euclidean metric of components  \cite{marco} $g_{ij} =2 [E- V(q)] \delta_{ij}$, 
whence the infinitesimal arc element $ds^2= 2[E- V(q)]^2 dq_i\ dq^i$; then Newton equations  are retrieved from the geodesic  equations
$$
\frac{d^2q^i}{ds^2} + \Gamma^i_{jk}\frac{d q^j}{ds}\frac{d q^k}{ds} =0\ ,
$$
 where  $\Gamma^i_{jk}$ are the Christoffel connection coefficients of the manifold.
The degree of instability of the dynamics is described by means of the Jacobi--Levi-Civita equation for the
geodesic spread  
$$
\frac{\nabla^2 J}{ds^2} + R(J, \dot\gamma )\dot\gamma =0\ ,
$$
 where the vector field $J$  locally measures the
distance between nearby geodesics,  $\frac{\nabla}{ds}$ is the covariant derivative along the configuration space geodesic ${\dot\gamma}$, and $R(\cdot,\cdot)$ is the Riemann curvature tensor. The  largest Lyapunov exponent for high dimensional Hamiltonian flows is found to depend on the curvature ``landscape'' of the configuration space manifold \cite{book}.  Hence, a natural consequence has been to investigate whether the occurrence of phase transitions has some peculiar counterpart in geometrical changes of the manifolds underlying the flows. And it has been discovered that this is actually the case.
Moreover, the peculiar geometrical changes associated with phase transitions were discovered to be the effects of deeper topological changes of the potential level sets $\Sigma_v^{V_N}: =\{ V_{N}(q_1,\dots,q_N) = v \in{\mathbb R}\}$ in configurations space, and, equivalently, of the balls $\{M_{v}^{V_N}=V_N^{-1}((-\infty,v])\}_{v \in{\mathbb R}}$ bounded by the $\Sigma_v^{V_N}$. 

A topological approach to the study of phase transitions has been considered for a variety of systems, ranging from those
undergoing entropy driven transitions \cite{carlsson1,barish} (having also applications to robotics), and hard spheres systems \cite{mason},
to quantum phase transitions \cite{brody,BFS,volovik}, glasses and supercooled liquids
\cite{angelani,stillinger}, classical models in statistical mechanics \cite{risau,schilling,fernando}, discrete spin models \cite{cimasoni}, DNA denaturation \cite{grinza}, peptide structure \cite{becker}, to quote just a few of them. 
In fact, in many contexts, well before an explicit formulation of the TH \cite{CCCP,CCCPPG}, topological concepts were implicitly entering the study of phase transitions while talking of energy landscapes \cite{brooks,wales} and of saddle points for disordered systems, glasses \cite{angelani,stillinger}, spin glasses: saddle points being critical points in the language of Morse theory of differential topology.  

On a completely different field, more recently, handling Big Data - outsourcing from complex systems - through methods referred to as Topological Data Analysis (TDA) it  happens to highlight the existence of phase transition-like phenomena in the absence of a statistical measure. Here the concept of phase transition is intended as the emergence of qualitatively new properties when a control parameter crosses a critical value (the prototype can be dated back to the Erd\"os-Renyi giant component appearance in random graphs). To quote a fascinating example in this field, in Ref.\cite{brain} the discovery is reported of topological phase transitions in functional brain networks by merging concepts from TDA, topology, geometry, physics, and network theory.
 
The present paper is organized as follows: in Section II the basic definitions and concepts of extrinsic geometry of hypersurfaces is recalled for the sake of self-containedness, and the definition of asymptotic diffeomorphicity is also therein introduced. In Section III the Main theorem is formulated and proved; this theorem  states that a topological change of the potential level sets of a physical system is a necessary condition for the appearance of a phase transition. Finally, in Section IV the problem raised by the counterexample to a preceding formulation of the Main theorem is fixed. Section V is devoted to some concluding remarks, two appendices contain computational details, and a third appendix addresses some past controversial points.
 
\section{Topological origin of phase transitions}
\smallskip

On the one side the study of the Hamiltonian dynamical counterpart of phase transitions, combined with the geometrization of Hamiltonian dynamics, has led to
find out the crucial role of topology at the grounds of these transition phenomena, on the other side a mathematical relationship exists between macroscopic thermodynamics and topological properties of the manifolds $M_{v}^{V_N}$, as expressed by \cite{book}
\begin{equation}
S_N(v) =({k_B}/{N}) \log \left[ \int_{M_{v}^{V_N}}\ d^Nq\right] 
=\frac{k_B}{N} \log \left[ vol
[{M_{v}^{V_N}\setminus\bigcup_{i=1}^{{\cal N}(v)} \Gamma(x^{(i)}_c)}]\ +
\sum_{i=0}^N w_i\ \mu_i (M_{v}^{V_N})+ {\cal R} \right]  ,\label{exactS}
\end{equation}
where $S_N$ is the configurational entropy, $v$ is the potential energy,  and the
$\mu_i(M_{v}^{V_N})$ are the Morse indexes (in one-to-one correspondence
with topology) of the manifolds $M_{v}^{V_N}$; in square brackets: the first term is the result of the excision of certain neighborhoods of the critical points of the interaction potential from  $M_{v}^{V_N}$; the second term
is a weighed sum of the Morse indexes, and the third term is a smooth function of $N$ and $v$.

\begin{figure}[h!]
 \centering
 \includegraphics[scale=0.4,keepaspectratio=true,angle=-90]{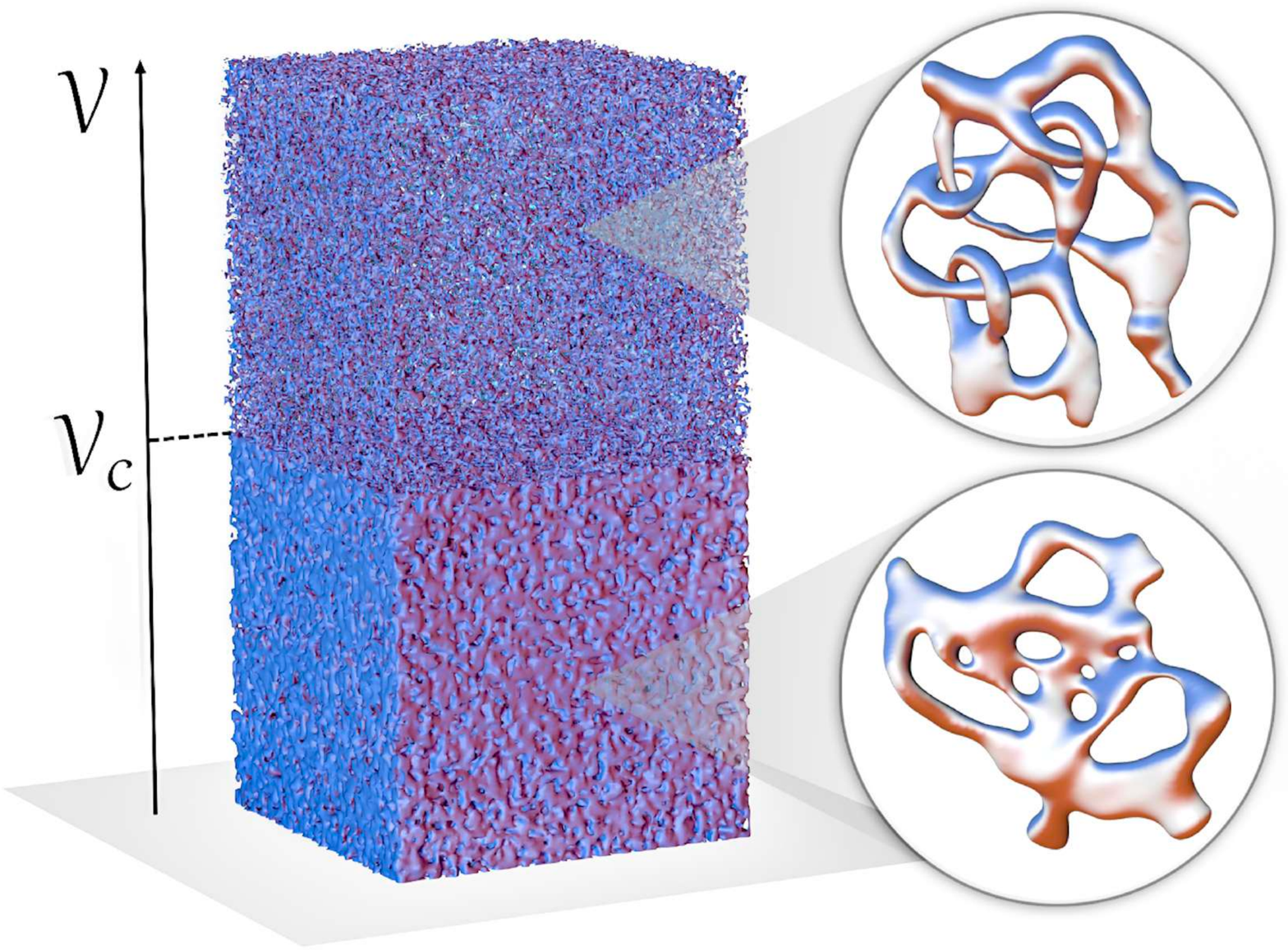}
 \vskip -0.3truecm
 \caption{Low-dimensional pictorial representation of the transition between complex topologies 
as a metaphor of the origin of a phase transition. From the ground level up to the 
crossover level $v_c$, the manifolds $M_v$ have a genus which increases with $v$.
Above the crossover level $v_c$, the manifolds $M_v$ have also a nonvanishing linking 
number which increases with $v$. }
\vskip 0.3truecm
\label{Cmplx}
\end{figure}

As a consequence, major topology changes with $v$ of the submanifolds $M_{v}^{V_N}$  - bringing about sharp changes
of the potential energy pattern of at least some of the $\mu_i (M_{v}^{V_N})$  - can affect the $v$-dependence of $S_N(v)$ and of its derivatives. 

Hence, it has been surmised \cite{book} that, at least for a broad class of physical systems, phase transitions stem from a suitable change of the topology of the potential level sets $\Sigma_v^{V_N}$ and, equivalently, of the manifolds $M_{v}^{V_N}$,  when $v$, playing the role of the control parameter, takes a critical value $v_c$. 
This hypothesis has turned into the start of a new theory by putting together several studies on specific models \cite{book,physrep} and  
two theorems  \cite{prl1,NPB1,NPB2}. These theorems state that an equilibrium phase transition - is {\it necessarily} due to appropriate
topological transitions in configuration space.
However, a counterexample to these theorems has been found in Ref.\cite{kastner} thus undermining this version of the topological theory of phase transitions. The counterexample is provided by the second order phase transition of the $2D$ lattice $\phi^4$-model that occurs at a critical value $v_c$ of the potential energy density which belongs to a broad interval of $v$-values void of critical points of the potential function. The difficulty raised by this counterexample has stimulated a deeper investigation of the transition of the 
$\phi^4$-model which led to figure out a crucial point associated with the breaking of the $\mathbb{Z}_2$ symmetry (and possibly with the breaking of discrete symmetries in general), that is, the possibility of a phase transition to stem from an asymptotic loss of diffeomorphicity of the relevant manifolds \cite{vaff}. 
In what follows this fact is formalized into a new and more consistent version of the theory. Other alleged difficulties of the theory are briefly discussed in  Appendix C.

\subsection{Basic definitions and concepts} 
Consider now an open set of $v$-values $I\subseteq \mathbb{R}$ such that the cylindrical subset of configuration space $\Gamma_I^N=\bigcup_{{v}\in {I}}\Sigma^{V_N}_{v}$
contains only non-singular level sets, that is,  $\nabla V_N(q)\neq 0$ for any $q\in\Gamma^N_{I}$, meaning that $V_N$ has no critical points for any ${v}\in {I}$.

 For any  ${v}_0,{v}_1\in{I}$,  the two level sets $\Sigma^{V_N}_{v_0}\subset \Gamma^N_{{I}}$ and 
 $\Sigma^{V_N}_{v_1} \subset \Gamma^N_{{I}}$ are diffeomorphic 
under the action of an explicitly known diffeomorphism given by the integral lines of the vector field  $\boldsymbol{\xi}_N={\nabla{V}_N}/{\|\nabla{V}_N\|^2}$, that is, any initial condition $\textit{\textbf{q}}_\alpha\in\Sigma^{V_N}_{v_0}$ is diffeomorphically mapped onto a point $\textit{\textbf{q}}_\beta\in\Sigma^{V_N}_{v_1}$ by the equation \cite{hirsch}
\begin{equation}\label{Hirsch-eq}
\frac{d{\textit{\textbf{q}}}}{dv} =\dfrac{\nabla{V}_N}{\|\nabla{V}_N\|^2} \qquad .
\end{equation} 

\begin{figure}[h]
\includegraphics[scale=0.32,keepaspectratio=true,angle=-90]{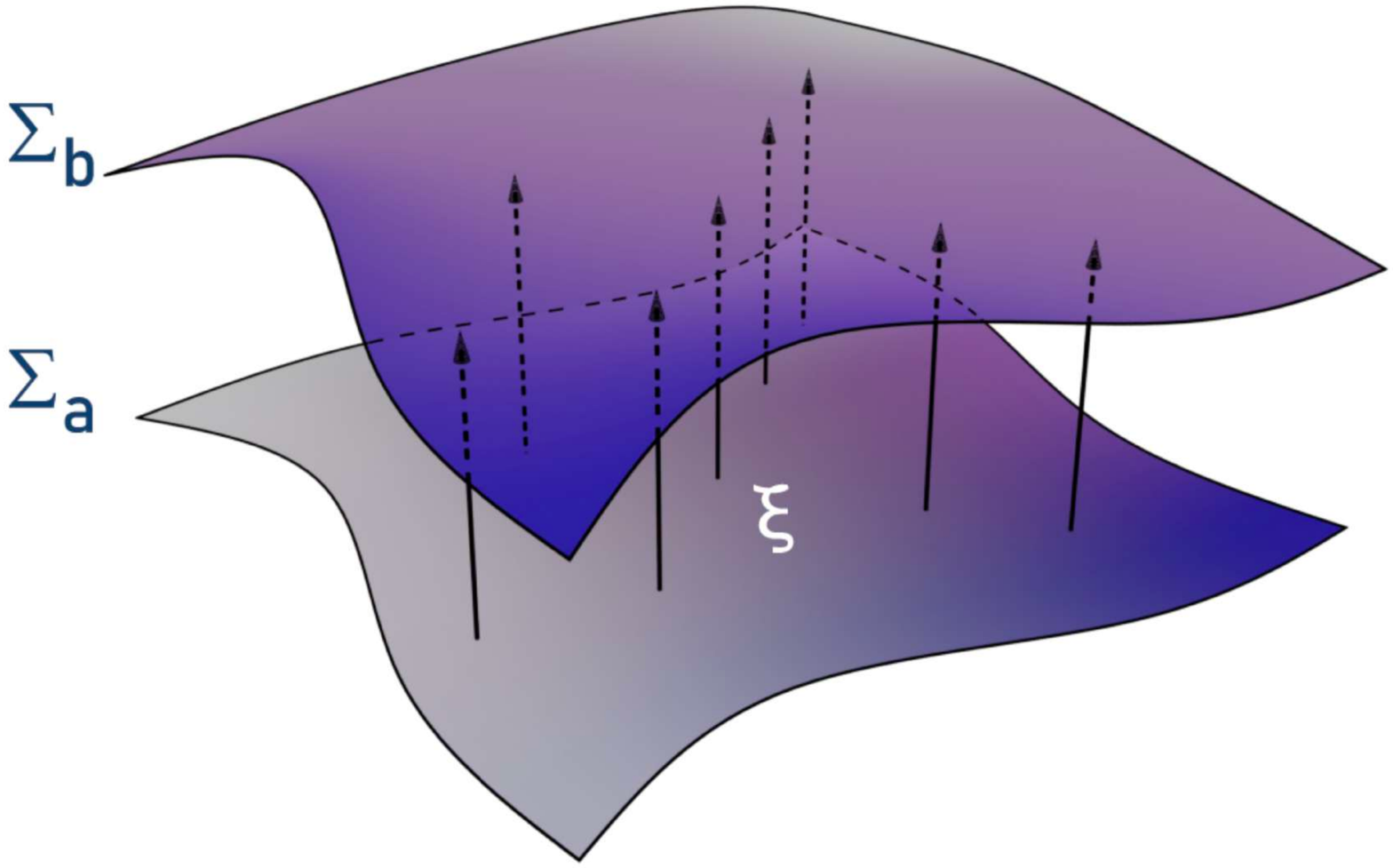}
 \caption{Pictorial representation of the action of the vector field $\boldsymbol{\xi}$ in Eq.\eqref{Hirsch-eq} diffeomorphically mapping each point of the level set $\Sigma_a$ onto each point of the level set $\Sigma_b$.}
\label{Hirsch-fig}
\end{figure}  

\subsection{Extrinsic geometry of hypersurfaces}
In this section some basic definitions and concepts are given about the
extrinsic geometry of hypersurfaces of a Euclidean space. The basic tool consists in measuring the way 
of changing from point to point on the surface of the normal direction in order to describe how the 
$n$-surface $\Sigma$ curves around in 
${\mathbb R}^N$. The rate of change of the normal vector ${\cal N}$ at a point $x\in\Sigma$ in a given 
direction $\textit{\textbf{u}}$ is described by the {\it shape operator} (also known as Weingarten's map)
$L_x(\textit{\textbf{u}}) = - \nabla_{\textit{\textbf{u}}}\ {\cal N}= - ({\textit{\textbf{u}}}\cdot\nabla){\cal N}$, 
where
$\textit{\textbf{u}}$ is a tangent vector at $x$ and $\nabla_{\textit{\textbf{u}}}$ is the directional
derivative; gradients and vectors are represented in ${\mathbb R}^N$.
\begin{figure}
\begin{center}
\includegraphics[scale=0.3,keepaspectratio=true,angle=-90]{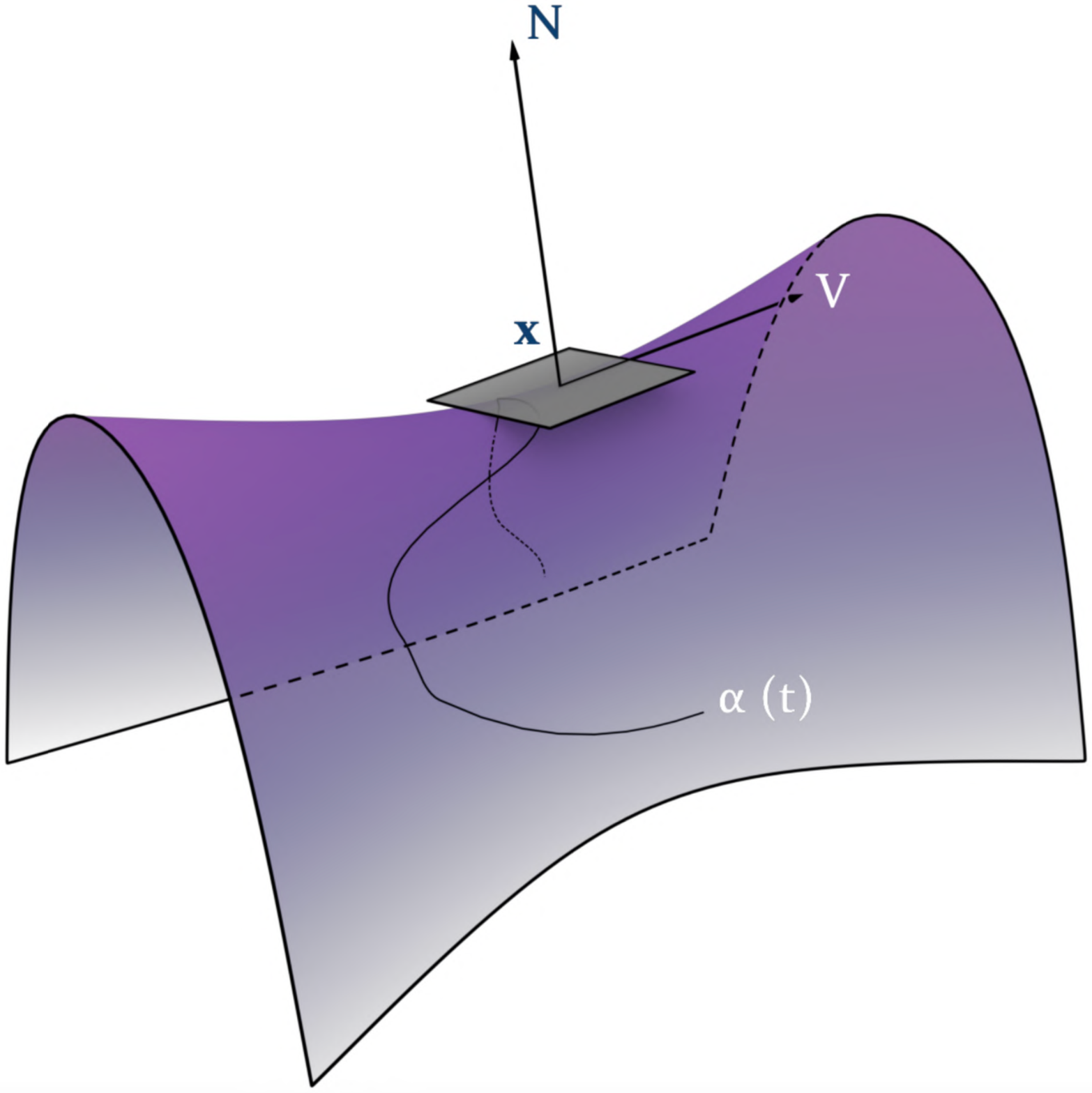}
\end{center}
\caption{Illustration of the items entering the construction of the shape operator of a
surface.  }
\label{wein-map}
\end{figure}

The constant-energy hypersurfaces in the phase space of Hamiltonian systems or of the 
equipotential hypersurfaces in configuration space, are the level sets of regular functions
and, for the level sets defined via a regular real-valued function $f$ as 
$\Sigma_a:=f^{-1}(a)$, the normal vector is ${\cal N}= \nabla f/\Vert\nabla f\Vert$.
Let $\{{\bf e}_\mu\}_{\mu =1,\dots,N}
=\{{{\bf e}_1,\dots,{\bf e}_n},{\cal N}\}$, with
${\bf e}_\alpha\cdot{\bf e}_\beta =\delta_{\alpha,\beta}$ and denote  with 
Greek subscripts, $\alpha =1,\dots,N$, the components in  the embedding space
${\mathbb R}^N$, and with Latin subscripts, $i =1,\dots,n$, the components on a
generic tangent space $T_x\Sigma_a$ at $x\in\Sigma_a$.
We consider  the case of codimension one, that is, $N=n+1$.

From $\partial_\mu {\cal N}_\alpha {\cal N}_\alpha =0= 2{\cal N}_\alpha \partial_\mu {\cal N}_\alpha$ we
see that for any $\textit{\textbf{u}}$,  we have 
${\cal N}\cdot L_x(\textit{\textbf{u}})= - {\cal N}_\alpha \textit{\textbf{u}}_\mu \partial_\mu {\cal N}_\alpha =0$, which
means that $L_x(\textit{\textbf{u}})$ projects on the tangent space $T_x\Sigma_a$.

Now the {\it principal curvatures} $\kappa_1,\dots,\kappa_n$ of $\Sigma_a$
at $x$ are the eigenvalues of the shape operator restricted to $T_x\Sigma_a$.
Considering the matrix ${\cal L}_x$ to be the restriction of $L_x$ to $T_x\Sigma_a$
\[
{\cal L}_{ij}(x) = {\bf e}_i\cdot L_x({\bf e}_j) = - ({\bf e}_i)_\alpha
({\bf e}_j)_\beta \partial_\beta {\cal N}_\alpha\ ,
\]
then the {\it mean curvature} is defined as
\begin{equation}
H(x)= \frac{1}{n}{\rm Tr}^{(n)}{\cal L}_{ij}(x)=
\frac{1}{n}\sum_{i=1}^n\kappa_i \ .
\label{meancurvat}
\end{equation}
The computation of the analytic expression of the mean curvature $H$ proceeds from
\begin{equation}
H(x)= \frac{1}{n}{\rm Tr}^{(n)}{\cal L}_{ij}(x)=
- \frac{1}{n}\sum_{i=1}^n ({\bf e}_i)_\alpha ({\bf e}_i)_\beta \partial_\beta
{\cal N}_\alpha~.
\label{meancurvatur}
\end{equation}
Defining $A_{\mu \nu}=({\bf e}_\mu)_\nu$, so that $A A^T={\mathbb I}$, we have
\[
\sum_{i=1}^n ({\bf e}_i)_\alpha ({\bf e}_i)_\beta =\delta_{\alpha\beta} -
{\cal N}_\alpha {\cal N}_\beta
\]
and thus
\begin{eqnarray}
H(x)= -\frac{1}{n}(\delta_{\alpha\beta} - {\cal N}_\alpha {\cal N}_\beta)\partial_\beta
{\cal N}_\alpha =- \frac{1}{n}\partial_\alpha {\cal N}_\alpha = - \frac{1}{n}\nabla \cdot\left(
\frac{\nabla f}{\Vert\nabla f\Vert}\right)~.
\label{M1}
\end{eqnarray}

\subsection{Asymptotic diffeomorphicity}
In Ref.\cite{vaff} it has been numerically found that the phase transition undergone by the $2D$ lattice $\phi^4$ model actually corresponds to a major topological change of the potential level sets of the model, also in absence of critical points of the potential. This topological change corresponds to an asymptotic breaking of the topological transitivity of the potential level sets, what can be formalised as an asymptotic loss of diffeomorphicity of the same manifolds in the broken symmetry phase. Hence a crucial hint to fix the problem stemming from the counterexample given by the $\phi^4$ model that has been hitherto considered fatal.

The first step to fix the problem thus consists in  defining asymptotic diffeomorphicity,
what is easily done by observing that a vector valued function of several variables, $f:\mathbb{R}^n\rightarrow\mathbb{R}^n$, is of differentiability class ${\cal{C}}^{l}$ if all the partial derivatives $(\partial^lf/\partial x_{i_1}^{l_1}\dots\partial x_{i_k}^{l_k})$ exist and are continuous, where each of $i_1,\dots,i_k$ is an integer between $1$ and $n$ and each $l_1,\dots,l_k$ is an integer between $0$ and $l$, and $l_1+\dots +l_k=l$. Then, by taking advantage of the explicit analytic representation of the vector field generating the diffeomorphism $\boldsymbol\xi_N:\Gamma_{I}^{N}\rightarrow T\Gamma_{I}^{N}$ previously given, uniform convergence in $N$ of the sequence $\{\boldsymbol\xi_N\}_{N\in\mathbb{N}}$ - and thus asymptotic diffeomorphicity in some class ${\cal{C}}^{l}$ - can be defined after the introduction of an appropriate norm containing all the derivatives up to $(\partial^l \boldsymbol\xi_N/\partial q_{i_1}^{l_1}\dots\partial q_{i_k}^{l_k})$.
At any fixed $N\in \mathbb{N}$,  in the absence of critical points of a confining potential $V_N$, the level sets $\Sigma_v^{V_N}$ are non singular $(N-1)$-dimensional hypersurfaces in $\mathbb{R}^N$.
Let us consider the already defined cylindrical subset of configuration space $\Gamma_I^N=\bigcup_{{v}\in {I}}\Sigma^{V_N}_{v}$ containing only non-singular level sets.

The lack of asymptotic breaking of diffeomorphicity is defined by  introducing a norm for the
$\boldsymbol{\xi}_N$ that allows to compare the diffeomorphisms at different dimensions
\begin{equation}\label{normaxi}
\|\boldsymbol{\xi}_N \|_{C^k(\Gamma^N_{{I}_0})}=\sup_{\textit{{q}}_0\in\Gamma^N_{{I}_0}}\|\boldsymbol{\xi}_N\|+
\sum_{l=1}^{k}\sum_{\{ i_k\}}\sum_{j=1}^N\|{\nabla^l_{\{ i_k\}}}{\xi}_j\|_{\Gamma^N_{{I}_0}}
\end{equation}
where $\{ i_k\}$ stands for a multi-index and $\|{\nabla^l_{\{ i_k\}}}{\xi}_j\|_{\Gamma^N_{{I}_0}}$ is the norm of the $l$-th differential operator with $l_1+\dots +l_k=l$
\begin{equation}
\|{\nabla^l_{\{ i_k\}}}{\xi}_j\|_{\Gamma^N_{{I}_0}}=\sup_{\textit{{q}}_0\in\Gamma^N_{{I}_0}}\left\vert \dfrac{\partial^l{\xi}_j}{\partial q_{i_1}^{l_1}\dots\partial q_{i_k}^{l_k}}\right\vert \  .
\label{norma}
\end{equation}
The sequence of families of manifolds  $\left\{\Gamma^N_{{I}_0}\right\}_{N\in\mathbb{N}}$ is said to
asymptotically preserve the $C^{k}$-diffeomorphicity  among the hypersurfaces  $\Sigma_v^{V_N}$ - foliating each family - if there exists 
$B\in\mathbb{R}^{+}$ such that
\begin{equation}
\label{eq:AsympDiffCk}
\|\boldsymbol{\xi}_N \|_{C^k\left(\Gamma^N_{{I}_0}\right)}\leq B<+\infty \qquad \forall N\in \mathbb{N}.
\end{equation}
As a consequence, from this condition we get  
$\|\nabla{V}_N\|=\|{\boldsymbol\xi}_N\|^{-1}\geq 1/B=C>0$ for each $\textit{q}_0\in\Gamma_{{I}_0}^N$ and all $N\in\mathbb{N}$, ruling out the existence of asymptotic critical points (i.e. $\|\nabla{V}_N\|\rightarrow 0$ for $N\rightarrow\infty$).

The analytic condition \eqref{eq:AsympDiffCk} entails remarkable consequences on the extrinsic geometry of the potential level sets. In fact, 
using $\sum_i \Vert X_i\Vert \ge \Vert \sum_i X_i\Vert$, from Eq. \eqref{norma} at the lowest order with the aid of  a normalised vector $\textit{\textbf{u}}$ tangent at $\textit{q}_0$ to a 
$\Sigma_v^n\subset \Gamma^N_{{I}_0}$, that is, $\textit{\textbf{u}}\in T_{\textit{q}_0}\Sigma_v^n$, we can build the quadratic forms
\begin{equation}\label{kappa}
\sum_{i,j=1}^N\Vert (\partial_i\xi_j) u_i u_j \Vert \ge \left\Vert \sum_{i,j=1}^N  \left(\partial_i \dfrac{\partial_j V_N}{\Vert\nabla V_N\Vert^2}\right) u_i u_j \right\Vert
\end{equation}
where $\partial_i=\partial/\partial q^i$.
 With implicit summation on repeated indices the r.h.s. of Eq.\eqref{kappa} is rewritten as
\begin{eqnarray}\label{Qform}
 && \left\Vert \left[\dfrac{1}{\Vert\nabla V_N\Vert}\left(\partial_i \dfrac{\partial_j V_N}{\Vert\nabla V_N\Vert}\right) 
 +  \dfrac{\partial_j V_N}{\Vert\nabla V_N\Vert}\partial_i \left( \dfrac{1}{\Vert\nabla V_N\Vert}\right)\right] u_i u_j  \right\Vert \nonumber \\
 &=& \left\Vert \dfrac{1}{\Vert\nabla V_N\Vert}\left(\partial_i \dfrac{\partial_j V_N}{\Vert\nabla V_N\Vert}\right) u_i u_j \right\Vert 
\end{eqnarray}
where the orthogonality, at any  point $\textit{q}_0$, between the vectors $\textit{\textbf{u}}$ and  

\noindent ${\cal N}=( \partial_1 V_N/\Vert\nabla V_N \Vert,\dots,\partial_N V_N/\Vert\nabla V_N\Vert )$  tangent and normal to $\Sigma_v^{V_N}$, respectively, has been used.
Through the shape operator (Weingarten map) of $\Sigma_v^{V_N}$ \cite{thorpe} at $\textit{q}_0$
\begin{equation}
L_{\textit{q}_0}(\textit{\textbf{u}}) = - L_{\textit{\textbf{u}}} {\cal N} = - (\nabla {\cal N}_1\cdot \textit{\textbf{u}}\ ,\dots,\nabla {\cal N}_N\cdot \textit{\textbf{u}})
\label{shape}
\end{equation}
the quadratic form $\kappa(\textit{\textbf{u}},\textit{q}_0) = \langle \textit{\textbf{u}}, L_{\textit{q}_0}(\textit{\textbf{u}})\rangle$ is found to coincide with the one given in Eq.\eqref{Qform} (last term). The quantity 
$\kappa(\textit{\textbf{u}},\textit{q}_0)$ is known as the \textit{normal curvature} of the level set $\Sigma_v^{V_N}$ at $\textit{q}_0$. 
Let $\{ \kappa_1(\textit{q}_0),\dots,\kappa_N(\textit{q}_0)\}$ denote the principal curvatures of  $\Sigma_v^{V_N}$ at $\textit{q}_0$, with the corresponding orthogonal principal curvature directions 
$ \{ \textit{\textbf{v}}_1,\dots,\textit{\textbf{v}}_N\}$,
then the normal curvature in the direction $\textit{\textbf{u}}\in T_{\textit{q}_0}\Sigma_v^N$  is given by
\begin{equation}
\kappa(\textit{\textbf{u}},\textit{q}_0)=\sum_{i=1}^N \kappa_i(\textit{q}_0)\langle \textit{\textbf{u}}, \textit{\textbf{v}}_i\rangle = \sum_{i=1}^N \kappa_i(\textit{q}_0)\cos^2\theta_i
\end{equation}
By choosing $\tilde{\textit{\textbf{u}}}\in T_{\textit{q}_0}\Sigma_v^N$ such that $\Vert\tilde{\textit{\textbf{u}}} \Vert =1$ and all the angles $\theta_i$ between $\tilde{\textit{\textbf{u}}}$ and $\textit{\textbf{v}}_i$ are equal to some $\tilde\theta$, we get
\begin{equation}\label{kappatilde}
\kappa(\tilde{\textit{\textbf{u}}},\textit{q}_0)=(\cos^2{\tilde\theta})\ \sum_{i=1}^N \kappa_i(\textit{q}_0) = (\cos^2{\tilde\theta})\ N\  H (\textit{q}_0)
\end{equation}
where $H (\textit{q}_0)$ is the mean curvature (the trace of the Weingarten map) at $\textit{q}_0$. Thus from Eqs.\eqref{Qform},\eqref{kappatilde} and \eqref{eq:AsympDiffCk} 
\begin{equation}
\left\Vert \dfrac{1}{\Vert\nabla V_N\Vert} (\cos^2{\tilde\theta})\ N\, H (\textit{q}) \right\Vert \le B <+\infty \qquad \forall N\in \mathbb{N}
\end{equation}
everywhere on $\Sigma_v^{V_N}$. Since $\Vert\nabla V_N\Vert\sim{\cal{O}}(N^{1/2})$ it follows that $H (\textit{q})\sim{\cal{O}}(N^{-1/2})$ everywhere on $\Sigma_v^{V_N}$ and uniformly in $N$. Therefore, the first remarkable consequence of asymptotic diffeomorphicity among the potential level sets is that their mean curvature
\begin{equation}
H (\textit{q})= \frac{1}{N}\sum_{i=1}^N\kappa_i(\textit{q})
\end{equation}
is everywhere uniformly bounded in $N$. However, this does not ensure the boundedness of each principal curvature (whose sign is not definite). A-priori two or more principal curvatures of the same value but of opposite sign could diverge and mutually compensate leaving $H (\textit{q})$ finite.
In order to get this missing information about the asymptotic boundedness of all the principal curvatures, let us consider the scalar curvature ${\mathscr R}$ of a level set $ V(\textit{q}) = v$, embedded in an Euclidean space of arbitrary dimension, which reads  \cite{Zhou}
\begin{equation}
{\mathscr R}(\textit{q}) = \frac{1}{N(N-1)}\sum_{i\leq j}^{1\dots N}\kappa_i(\textit{q})\kappa_j(\textit{q}) =\frac{1}{N(N-1)}\left\{ -\triangle\log\Vert\nabla V_N(\textit{q})\Vert + \nabla\cdot \left[ \triangle V_N(\textit{q}) \frac{\nabla V_N(\textit{q}) }{\Vert\nabla V_N(\textit{q})\Vert^2} \right] \right\}
\label{scalar0}
\end{equation}
let us notice that ${\mathscr R}$ is singular at the critical points of the potential, where $\nabla V_N(\textit{q})=0$, and can be arbitrarily large in their neighborhoods; then, using $\Vert\boldsymbol{\xi}\Vert = \Vert\nabla V_N(\textit{q})\Vert^{-1} $, this can be rewritten as
\begin{equation}
{\mathscr R} = \frac{1}{N(N-1)}\left\{ -\triangle\log \frac{1}{\Vert\boldsymbol{\xi}\Vert } + \nabla\cdot \left[ \triangle V_N(\textit{q})\  \boldsymbol{\xi}   \right] \right\} \ ,    
\label{scalar1}
\end{equation}
then, trivial computations (sketched in Appendix A) of the r.h.s. of this equation under the assumption of asymptotic diffeomorphicity [Eqs.\eqref{normaxi},\eqref{norma} and \eqref{eq:AsympDiffCk}] yield uniform boundedness also of ${\mathscr R}(\textit{q})$ entailing uniform boundedness in $N$ of each $\kappa_i(\textit{q})$ everywhere on each potential level set.
\begin{figure}[h!]
\includegraphics[scale=0.3,keepaspectratio=true,angle=0]{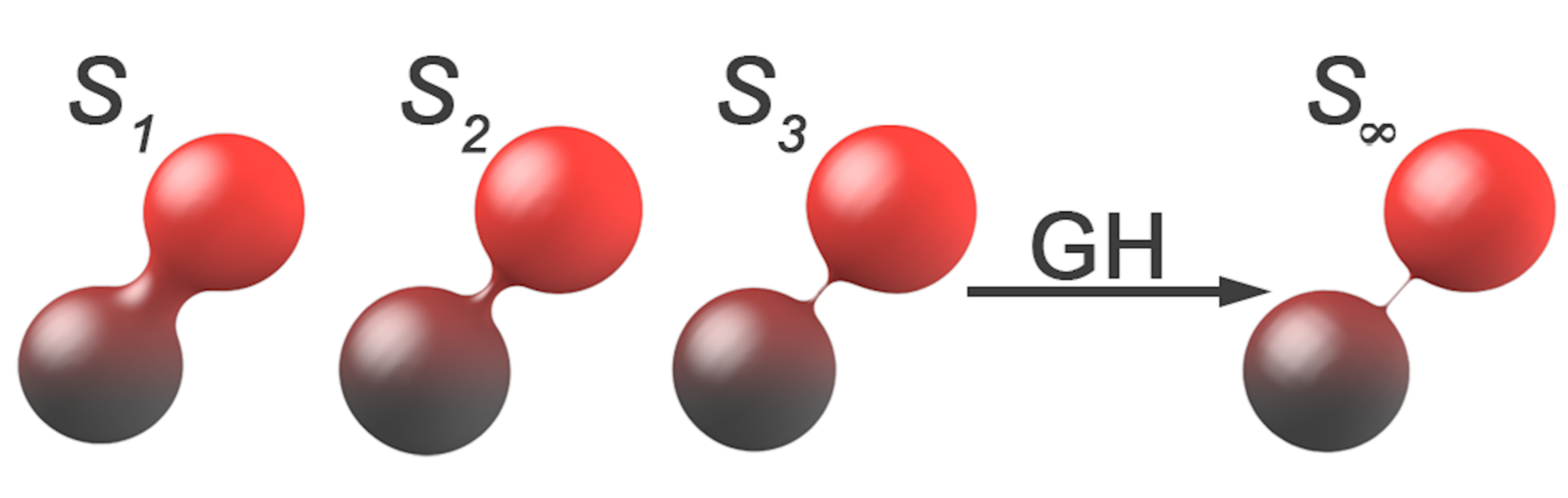}
 \caption{Sequence of diffeomorphic manifolds (of the same dimension) with a limit manifold which is not diffeomorphic to the  members of the sequence. The infinitely tiny bridge between the two spheres of $S_\infty$ has infinite mean curvature.}
\label{GHlimt}
\end{figure}  

To help intuition to get a hold of the relationship between boundedness of mean and scalar curvatures and asymptotic diffeomorphicity, we qualitatively illustrate in Figure \ref{GHlimt}  the opposite situation, known as Gromov-Hausdorff limit \cite{sormani}, where a sequence of diffeomorphic manifolds of fixed dimension have a limit manifold which is not diffeomorphic to the other members of the sequence. The handles of these dumbbell shaped manifolds shrink to an asymptotic infinitely tiny cylinder of vanishing radius and thus of diverging transversal principal curvature, that is, of divergent mean curvature.  

 \begin{remark}. Summarizing, the assumption of asymptotic diffeomorphicity  means that, for any pair of densities $\vb$ and $\vb^\prime$ in some assigned interval $I_{\vb}=[\vb_0, \vb_1]$ and $N$ arbitrarily large,  the corresponding manifolds $\Sigma_{N\vb}^{V_N}$ are diffeomorphic under the action of the diffeomorphism-generating vector fields $\boldsymbol{\xi}_{N_k}$
 \begin{eqnarray}
&& \Sigma_{N_1\vb}^{V_{N_1}}\quad \overset{\boldsymbol{\xi}_{N_1}}\longrightarrow \quad  \Sigma_{N_1\vb^\prime}^{V_{N_1}}\nonumber\\
 &&\Sigma_{N_2\vb}^{V_{N_2}}\quad \overset{\boldsymbol{\xi}_{N_2}}\longrightarrow \quad  \Sigma_{N_1\vb^\prime}^{V_{N_2}}\nonumber\\
 &&\vdots \quad\quad\quad\quad\quad\quad\quad\quad\quad\quad\quad\quad \vb, \vb^\prime\in[\vb_0,\vb_1 ],\quad k\in{\mathbb N}\\
&&  \Sigma_{N_k\vb}^{V_{N_k}}\quad \overset{\boldsymbol{\xi}_{N_k}}\longrightarrow \quad  \Sigma_{N_k\vb^\prime}^{V_{N_k}}\nonumber\\
 &&\vdots\nonumber
 \end{eqnarray}
provided that the norm of the vector fields $\boldsymbol{\xi}_{N_k}$ is uniformly bounded according to Eq.\eqref{normaxi}. Under this condition, all the principal curvatures $\kappa_i(\textit{q})$ of every manifold in the above diagram are uniformly bounded with $N$. Moreover, after the {\rm Non-critical neck Theorem \cite{palais} } all the above manifolds $\Sigma_{N_k\vb}^{V_{N_k}}$ for any $\vb\in[\vb_0,\vb_1]$ are free of critical points of the potential functions $V_N$, that is of points where $\nabla V_N=0$.
 \end{remark}

\section{A necessity  theorem}
In its original formulation, given in Refs.\cite{prl1,NPB1}, the theorem establishing the necessary topological origin of a phase transition was lacking a fundamental hypothesis that has led to the paradoxical situation of being falsified \cite{kastner} through the example of a phase transition still related with a change of topology in configuration space, though asymptotic in the number of degrees of freedom \cite{vaff}, and in the absence of critical points of the potential. 

The missing hypothesis suggested by the study of Ref. \cite{vaff} is to require also asymptotic diffeomorphicity of the potential level sets in order to correspondingly get uniform convergence of the Helmholtz free energy in a differentiability class that rules out first and second order phase transitions.

\begin{remark}.The notation $N\in\mathbb{N}^\#$ means that $N\rightarrow\infty$ is included. \end{remark}
\medskip

\subsection*{\large Main Theorem}
\begin{theorem}[Absence of phase transitions  under diffeomorphicity] 
\noindent Let $V_N(q_1,\dots,q_N): {\mathbb R}^N \rightarrow{\mathbb R}$, be a smooth, nonsingular, finite-range
potential. Denote by $\Sigma_v^{V_N}:= V_N^{-1}(v)$, $v\in{\mathbb R}$, its
{\em level sets}, or {\em equipotential hypersurfaces}, in
configuration space.\\
Then let $\vb =v/N$ be the potential energy per degree of freedom.\\
If for any pair of values $\vb$ and $\vb^\prime$ belonging  to a
given interval $I_{\vb}=[\vb_0, \vb_1]$ and for any $N>N_0$ with $N\in\mathbb{N}^\#$ we have\\
\centerline{$\Sigma_{N\vb}^{V_N}\approx \Sigma_{N\vb^\prime}^{V_N}$\ , }
\vskip 0.2truecm
\noindent that is, $\Sigma_{N\vb}^{V_N}$ is {\em diffeomorphic} to
$\Sigma_{N\vb^\prime}^{V_N}$, including {\em asymptotically diffeomorphic}, then the sequence of the Helmholtz free
energies $\{ F_N (\beta)\}_{N\in{\mathbb N}}$---where $\beta =1/T$
($T$ is the temperature) and $\beta\in I_\beta =(\beta (\vb_0),
\beta (\vb_1))$---is {\em uniformly} convergent at least in
${\mathscr C}^2(I_\beta\subset{\mathbb R})$, so that $F_\infty \in{\mathscr C}^2(I_\beta\subset{\mathbb R})$
and neither first- nor second-order phase transitions can occur in
the (inverse) temperature interval $(\beta (\vb_0), \beta (\vb_1))$.
\end{theorem}

\begin{remark}. {\rm 
The configurational entropy $S_N(\vb)$ is
related to the configurational canonical free energy, $f_N$ in \eqref{freeEnergy}, for
any $N\in{\mathbb N}$, $\vb\in{\mathbb R}$, and $\beta \in{\mathbb R}$ through the Legendre
transform 
    \begin{eqnarray}
     - f_N(\beta) =  \beta \cdot \vb_N -  S_N(\vb_N)
    \label{legendre-tras}
    \end{eqnarray}
where the inverse of the configurational temperature $T(v)$  is given by $\beta_N(\vb)= {\partial S_N(\vb)}/{\partial \vb}$.
By following Ref.\cite{dualising}, let us consider the function $\phi(\vb)=f_N[\beta(\vb)]$, from $\phi^\prime(\vb) = -\vb\  [d \beta_N(\vb)/d\vb]$
it is evident that if $\beta_N(\vb)\in{\cal C}^k(\mathbb R)$ then also $\phi(\vb)\in{\cal C}^k(\mathbb R)$ and thus 
$S_N(\vb)\in{\cal C}^{k+1}(\mathbb R)$ while $f_N(\beta)\in{\cal C}^k(\mathbb R)$.   First and 
second order phase transitions are associated with a discontinuity in the first or second derivatives of $f_\infty(\beta)$, that is with $f_\infty(\beta)\in{\cal C}^0(\mathbb R)$ or $f_\infty(\beta)\in{\cal C}^1(\mathbb R)$, respectively. 
Hence a first order phase transition corresponds to a discontinuity of the second derivative of the 
entropy $ S_\infty(\vb)$, and a second order phase transition  corresponds to a discontinuity of the third derivative of the entropy $S_\infty(\vb)$.  }
\label{diffclassS}
\end{remark}

\begin{remark}. {\rm The proof of the Main Theorem  follows the same conceptual path given in Refs.\cite{prl1,NPB1}: 
a {\it topological change} of the equipotential hypersurfaces $\Sigma_v^{V_N}$ of configuration space is a {\it necessary} condition for the occurrence of 
a thermodynamic phase transition if we prove the {\it equivalent proposition} that if any two hypersurfaces $\Sigma_v^{V_N}$ and ${\Sigma_{v^\prime}}^{V_N}$ with
$v(N), v^\prime (N) \in (v_0(N),v_1(N))$ are {\it diffeomorphic} for all $N\in\mathbb{N}^\#$,  then {\it no phase transition} can occur in the (inverse) temperature interval
$[\lim_{N\rightarrow\infty}\beta (\vb_0(N)),\lim_{N\rightarrow\infty} \beta (\vb_1(N))]$.  }
\end{remark}

\noindent\textbf{Proof.}  
{\rm For standard Hamiltonian systems (i.e. quadratic in the momenta) the relevant information is carried by the configurational microcanonical ensemble, where the configurational canonical free energy is }
\begin{equation}\label{freeEnergy}
    f_N(\beta)\equiv f_N(\beta; V_N)=   \frac{1}{N} \log \int_{(\Lambda^d)^{\times n}}dq_1\dots dq_N\ \exp [-\beta V_N(q_1,\dots, q_N)]
\end{equation}
{\rm     with 
and the configurational microcanonical entropy  (in units s.t. $k_B=1$) is 
\begin{equation}
      S_N(\vb) \equiv S_N(\vb;V_N) =\frac{1}{N} \log{ \int_{(\Lambda^d)^{\times n}} dq_1\cdots dq_N\ \delta
[V_N(q_1,\dots, q_N) - v] ~\label{pallaM}} \, .\nonumber
\end{equation}
Then $S_N(\vb)$ is
related to the configurational canonical free energy, $f_N$, for
any $N\in{\mathbb N}$, $\vb\in{\mathbb R}$, and $\beta \in{\mathbb R}$ through the Legendre
transform in Eq.\eqref{legendre-tras}. }

From Lemma 1, proved after Lemmas 2 to 9, we have that in the limit $N\rightarrow\infty$
and at constant particle density, ${\rm vol} (\Lambda^d)^{\times n}/N\ =\ {\rm const}$, in the interval $I_{\vb}=[\vb_0, \vb_1]$
the sequence $\{S_N\}_{N\in{\mathbb N}^\#}$ is uniformly convergent in ${\mathscr C}^3(I_{\vb}\subset{\mathbb R})$ so that
$S_\infty\in{\mathscr C}^3(I_{\vb}\subset{\mathbb R})$ that is, the thermodynamic limit of the entropy is three times
differentiable, with continuous third-order derivative, in $I_{\vb}=[\vb_0, \vb_1]$. Hence in the interval $I_\beta=[\lim_{N\rightarrow\infty}\beta (\vb_0(N)),\lim_{N\rightarrow\infty} \beta (\vb_1(N))]$ the sequence of configurational free energies $\{ f_N(T)\}_{N\in\mathbb{N}^\#}$ is
{\it uniformly convergent} at least in ${\mathscr C}^2(I_\beta\subset{\mathbb R})$, so that we have 
\[
- f_\infty(\beta) =  \beta(\vb) \cdot \vb -  S_\infty(\vb)
\]
that is $\{ f_\infty(T)\}\in {\mathscr C}^2(I_\beta\subset{\mathbb R})$. 

Since a quadratic kinetic energy term of a standard Hamiltonian gives only a smooth contribution to the total Helmholtz free energy $F_N(\beta)$, also
the asymptotic function $F_\infty(\beta)$ has differentiability class ${\mathscr C}^2(I_\beta\subset{\mathbb R})$ so that we conclude that the corresponding
physical system does not undergo neither first- nor second-order phase transitions in the inverse-temperature interval $\beta\in I_\beta$. \ $\square$

\subsection*{\large Lemmas}

\begin{lemma}[Uniform upper bounds]
Let $V_N$ be a standard, short-range, stable, and confining
potential function bounded below. Let $\left \{ \Si_v^{V_N} \right
\}_{v\in{\mathbb R}}$ be the family of $(N-1)$-dimensional
equipotential hypersurfaces $\Si_v^{V_N}:=V_N^{-1}(v)$, $v\in{\mathbb R}$,
of ${\mathbb R}^N$. If
    \bena
    \forall N\in\mathbb{N}^\# ~~~and~\vb,\vb' \in {I_{\vb}}=[\vb_0,\vb_1],~~we~have~~
      \Si_{N \vb}^{V_N}~\approx~\Si_{N \vb'}^{V_N}\ , \nonumber
    \eena
then
    \bena
    \sup_{N,\vb\in I_{\vb}} \left\vert S_N({\vb})\right\vert
        < \infty~~~{\it and}~~~
    \sup_{N,\vb\in I_{\vb}} \left\vert \frac{\de^k
    S_N}{\de {\vb}^k}({\vb})\right\vert < \infty,~~k=1,2,3,4. \nonumber
    \eena
\label{derivees-majorees}
\end{lemma}
{\bf Proof.} The proof of this Lemma amounts to proving the Main Theorem and proceeds as follows. After Remark 2, the derivatives of the entropy are expressed in terms of the derivatives of the microcanonical configurational volume which, in turn, after Lemma 2 can be expressed  as surface integrals of functions of $\overline{\zeta}_N= {\rm div} (\overline{\boldsymbol{\xi}}_N)$ and its Lie derivatives, where $\overline{\boldsymbol{\xi}}_N$ is the vector field generating the diffeomorphisms among the specific potential energy level sets. Then these integrals are replaced by averages along Monte Carlo Markov Chains (MCMC) that can be defined to have as invariant measure the microcanonical configurational measure (Lemma 3 and Remark 3). After Lemmas 4 and 5, $\overline{\zeta}_N$ is proved to behave as a random gaussian process along the mentioned MCMCs, hence, after Remark 5 and Lemmas 6 to 9 the uniform bounds are derived of the derivatives of the entropy up to the fourth one. $\square$


\begin{lemma}[Derivation of integrals over regular level sets (\cite{federer}\cite{laurence})] 

Let $O \subset \mathbb{R}^p$ be a bounded open set. Let $\psi \in {\mathscr C}^{n+1} (\overline O)$
be constant on each connected component of the boundary $\partial O$ and $f \in {\mathscr C}^n (O)$.
Define $O_{t,t'}=\{x \in O\mid t<\psi (x) <t' \}$ and

\begin{equation}
F(v)= \int_{ \{ \psi=v \} } f~\D\sigma^{p-1}
\end{equation} 
where $d\sigma^{p-1}$ represents the Lebesgue measure of dimension $p-1$.
If $~C>0$ exists such that for any $x \in O_{t,t'}, \Vert \nabla
\psi (x) \Vert \geq C$, then for any $k$ such that  $0 \leq k \leq n$,
for any  $v \in ]t,t'[$, one has
\begin{equation}
\label{eq: FedLau_Formula}
\frac{\D^k F}{\D v^k}(v) =\int_{\{ \psi=v \}} A_{\psi}^k f~\D\sigma^{p-1} \ .
\end{equation}
with 
\begin{equation}
\label{eq: def_FedLauOperator}
A_{\psi} f = \nabla \left( \dfrac{\nabla \psi}{\|\nabla  \psi \| }f \right) \dfrac{1}{\| \nabla \psi \|}
\end{equation}
\label{Cor:Federer}
\end{lemma}

This Lemma allows to compute higher order derivatives of  the microcanonical volume $\Omega_{n}(\bv)$, and thus of the entropy, at any order by identifying $\psi$ with the potential $\overline{V}_N(\textit{\textbf{q}})= V(\textit{\textbf{q}})/N $. Let us introduce the following notations:  
$\overline{\zeta}_N= {\rm div} (\overline{\boldsymbol{\xi}}_N)$, 
\begin{equation}
\label{eq:chi_def}
\overline{\chi}_N=\|\overline{\boldsymbol{\xi}}_N\|=\dfrac{1}{\|\nabla\overline{V}_N\|}\,\, ,
\end{equation}
for the norm of $\overline{\boldsymbol{\xi}}_N$, and 
\begin{equation}
\label{eq:def_microcanonical_areaform}
\D\mu^{N-1}_{\bv}=\overline{\chi}_N \D\sigma_{\LevelSetfunc{\bv}{\overline{V}_N}}
\end{equation}
for the microcanonical area $(N-1)$-form of non critical energy level sets, and 
\begin{equation}
\mathcal{L}_{\overline{\boldsymbol{\xi}}}(\cdot)=(\overline{\boldsymbol{\xi}}\cdot\nabla)(\cdot)=\sum_{i=1}^{N}\dfrac{\partial^{i}\overline{V}}{\|\nabla \overline{V}\|^2}\partial_i(\cdot)
\end{equation}
for the Lie derivative along the flow of $\overline{\boldsymbol{\xi}}_N$. Then, given the microcanonical configurational volume
\begin{equation}
\Omega_N(\bv)=\intSigmaV{\bv}{N} \,\D\mu_{\bv}^{N-1}
\end{equation}
its derivatives are computed through the formula
\begin{equation}
\label{eq:def_recursiveDer_Omega_microcan}
\dfrac{\D^k \Omega_N}{\D \overline{v}^k}(\bv)=\intSigmaV{\bv}{N}\, \dfrac{1}{\overline{\chi}} A^{k}_{V}(\overline{\chi})\,\D\mu_{\bv}^{N-1}
\end{equation}
where $A^{k}_{V}(\overline{\chi})$ stands for a $k$-times repeated application of the operator
\begin{equation}
\label{eq:FedererOperator_microcan}
A_{V}(f) = f\overline{\zeta}_N+\mathcal{L}_{\overline{\boldsymbol{\xi}}_N}(f)\ . 
\end{equation}
\vspace{-0.3truecm}
\begin{remark}\textbf{(Derivatives of the entropy)}
The configurational microcanonical entropy density is given by
\begin{equation}
\label{eq: MicroCan_Entropy_xi}
\overline{S}_{N}(\overline{v})=\dfrac{1}{N}\log\Omega_{N}(\overline{v})=\dfrac{1}{N}\log\intSigmaV{\bv}{\overline{V}_N}\,\mathrm{d}\mu^{N-1}_{\overline{v}}
\end{equation}
and its derivatives are
\begin{equation}
\label{eq: microcanEntropy_Derivative}
\begin{split}
&\dfrac{\D \overline{S}_N}{\D
\overline{v}}(\overline{v})=\dfrac{1}{N}\dfrac{\Omega^{'}_N(\overline{v})}{\Omega_N(\overline{v})}\\
&\dfrac{\D^2 \overline{S}_N }{\D \overline{v}^2}(\overline{v})=\dfrac{1}{N}\,\left[
\dfrac{\Omega^{''}_N(\overline{v})}{\Omega_N(\overline{v})}-\left(\dfrac{\Omega^{'}_N(\overline{v})}{\Omega_N(\overline{v})}\right)^2 \right]\\
&\dfrac{\D^3 \overline{S}_N }{\D
\overline{v}^3}(\overline{v})=\dfrac{1}{N}\left[\dfrac{\Omega^{'''}_N(\overline{v})}{\Omega_N(\overline{v})}-3\dfrac{\Omega_N^{''}(\overline{v})}{\Omega_N(\overline{v})}
\dfrac{\Omega_N^{'}(\overline{v})}{\Omega_N(\overline{v})}+2\left(\dfrac{\Omega_N^{'}(\overline{v})}{\Omega_N(\overline{v})}\right)^3\right]\\
&\dfrac{\D^4 \overline{S}_N }{\D
\overline{v}^4}(\overline{v})=\dfrac{1}{N}\left[\dfrac{\Omega^{(iv)}_{N}(\overline{v})}{\Omega_N(\overline{v})}-4\dfrac{\Omega^{'''}_N(\overline{v})\Omega^{'}_N(\overline{v})}{\Omega_N^2(\overline{v})}+
12\dfrac{\Omega^{'2}_N(\overline{v})\Omega^{''}_N(\overline{v})}{\Omega_N^3(\overline{v})}-3\left(\dfrac{\Omega^{''}_N(\overline{v})}{\Omega_N(\overline{v})}\right)^2-
6\left(\dfrac{\Omega^{'}_N(\overline{v})}{\Omega_N(\overline{v})}\right)^4\right]\,.
\end{split}
\end{equation} 
where, after Lemma 2, the derivatives of configurational microcanonical volume $\Omega_N(\bv)$ up to the fourth order with respect to $\bv$  are found to be
\begin{equation}
\label{eq:Omega_Derivatives_Xi}
\begin{split}
&\dfrac{\D \Omega_N }{\D \bv}(\bv)=\intSigmaV{\bv}{\overline{V}_N}
\overline{\zeta}_N \,\D\mu^{N-1}_{\bv}\\
&\dfrac{\D^2 \Omega_N }{\D \bv^2}(\bv)=\intSigmaV{\bv}{\overline{V}_N}\left[
\overline{\zeta}_N^2+\mathcal{L}_{\overline{\boldsymbol{\xi}}_N}
(\overline{\zeta}_N)\right]\D\mu^{N-1}_{N\bv}\\
&\dfrac{\D^3 \Omega_N }{\D \bv^3}(\bv)=\intSigmaV{\bv}{\overline{V}_N}\left[
\overline{\zeta}_N^3+3\overline{\zeta}_N\mathcal{L}_{\overline{\boldsymbol{\xi}}_N}\left(\overline{\zeta}_N\right)+
\mathcal{L}^{(ii)}_{\overline{\boldsymbol{\xi}}_N}
(\overline{\zeta}_N)\right]\D\mu^{N-1}_{\bv}\\
&\dfrac{\D^4 \Omega_N
}{\D\overline{v}^4}(\overline{v})=\intSigmaV{\bv}{\overline{V}_N}\left[\overline{\zeta}_N^4+6\overline{\zeta}_N^2\mathcal{L}_{\overline{\boldsymbol{\xi}}_N}(\overline{\zeta}_N)+4\overline{\zeta}_N
\mathcal{L}^{(ii)}_{\overline{\xi}_N}(\overline{\zeta}_N)+3\left(\mathcal{L}_{\overline{\boldsymbol{\xi}}_N}(\overline{\zeta}_N)\right)^2+\mathcal{L}^{(iii)}_{\overline{\boldsymbol{\xi}}_N}(\overline{\zeta}_N)\right]\D\mu^{N-1}_{\bv}
\end{split}
\end{equation}

\end{remark}

On any $(N-1)$-dimensional hypersurface
$\Sigma_{N\vb}^{V_N}=V^{-1}_N(N\vb ) =\{X\in{\mathbb R}^{N}\ \vert \ V_N(X)
=N\vb\}$ of ${\mathbb R}^{N}$, we can define a homogeneous 
nonperiodic random Markov chain whose probability measure is the
configurational microcanonical measure \cite{book}, namely $d\sigma
/\Vert\nabla V_N\Vert$. We call this Markov chain a microcanonical-Monte Carlo Markov Chain (MCMC). In so doing, all the integrals giving configurational microcanonical averages are replaced by asymptotic averages along these MCMCs. Dropping the suffix $N$ of $V_N$ we have the following Lemma:

\begin{lemma} [Monte Carlo Markov Chains over regular level sets]

\label{mesure_ergodique} 
On each finite-dimensional level set
$\Sigma_{N\vb}=V^{-1}(N\vb )$ of a standard, smooth, confining,
short-range potential $V$ bounded below, and in the absence of
critical points, there exists a random Markov chain of points
$\{X_i\in{\Bbb R}^{N}\}_{i\in{\Bbb N_+}}$, constrained by the
condition $V(X_i) = N{\vb}$, which has
\begin{equation}
d\mu =\frac{d\sigma}{\Vert\nabla V\Vert}
\left(\int_{\Sigma_{N\vb}} \frac{d\sigma}{\Vert\nabla
V\Vert}\right)^{-1} \label{prob}
\end{equation}
as its probability measure, so that for a smooth function $F
:{\Bbb R}^{N}\rightarrow{\Bbb R}$ we have
\begin{equation}
 \left(\int_{\Sigma_{N\vb}}
\frac{d\sigma}{\Vert\nabla V\Vert}\right)^{-1}
\int_{\Sigma_{N\vb}}\frac{d\sigma}{\Vert\nabla V\Vert}\ F =
\lim_{n\rightarrow\infty}\frac{1}{n}\sum_{i=1}^n F(X_i)~.
\label{mcmcmc}
\end{equation}
\end{lemma}

{\bf Proof.} The level sets $\{\Sigma_{N\vb}\}_{\vb\in{\Bbb
R}}$ are compact hypersurfaces of ${\Bbb R}^{N}$, therefore
a partition of unity \cite{thorpe} can be defined on each hypersurface.
Then, let $\{U_i\}$, $1\leq i \leq m$ be an arbitrary finite
covering of $\Sigma_{N\vb}$ by means of domains of coordinates
(for example open balls), at any point of $\Sigma_{N\vb}$ an ensemble of smooth functions
$\{\varphi_i\}$ exists, such that $1\geq\varphi_i\geq 0$ and
$\sum_i\varphi_i =1$. 

By means of the partition of unity $\{\varphi_i\}$ on $\Sigma_{N\vb}$,
associated to a collection $\{U_i\}$ of one-to-one local
parametrizations of the compact and oriented hypersurfaces $\Sigma_{N\vb}$, 
the integral of a given smooth $(N-1)$-form $\omega$ is given by:
\[
\int_{\Sigma_{N\vb}} \omega^{(N-1)} =
\int_{\Sigma_{N\vb}}\left(\sum_{i =1}^m\varphi_i (x)\right)
\omega^{(N-1)}(x)= \sum_{i =1}^m \int_{U_i}
\varphi_i\omega^{(N-1)}(x)~.
\]
The existence of a Monte Carlo Markov
chain (MCMC) of assigned probability measure (\ref{prob}) on a given $\Sigma_{N\vb}$ is
constructively proved as follows.
Let us consider sequences of random values $\{x_i : i\in\Lambda\}$,
where $\Lambda$ is the finite set of indexes of the elements of the
partition of unity on $\Sigma_{N\vb}$, and where $x_i =(x^1_i,
\dots,x^{N-1}_i)$ are local coordinates with respect to $U_i$ of
an arbitrary representative point of the set $U_i$ itself. 
The weight $\pi (i)$ of the $i$th element of the partition is then defined
by
\begin{equation}
\pi (i)=\left( \sum_{k=1}^m \int_{U_k}\varphi_k\
\frac{d\sigma}{\Vert\nabla V\Vert} \right)^{-1}
\int_{U_i}\varphi_i\ \frac{d\sigma}{\Vert\nabla V\Vert}
\label{peso}
\end{equation}
and the transition matrix elements \cite{mcmc} are given by
\begin{equation}
p_{ij} = \min \left[ 1, \frac{\pi (j)}{\pi (i)}\right] \label{pij}
\end{equation}
satisfyinng the detailed balance equation $\pi (i) p_{ij}= \pi
(j) p_{ji}$. 
A random Markov chain $\{i_0, i_1\dots, i_k, \dots\}$ of
indexes induces a random Markov chain  of corresponding 
elements of the partition, that is of points 
$\{x_{i_0},x_{i_1},\dots, x_{i_k}, \dots\}$ on the hypersurface
$\Sigma_{N\vb}$.
Denote by $(x^1_P, \dots,x^{N-1}_P)$  the local
coordinates of a point $P$ on $\Sigma_{N\vb}$ and define a local
reference frame as $\{\partial/\partial
x^1_P,\dots,\partial/\partial x^{N-1}_P, n(P)\}$, with $n(P)$ 
the outward unit normal vector at $P$; by means of the matrix that operates a point-dependent 
change from this reference frame 
to the canonical basis $\{e_1,\dots,e_{N}\}$ of ${\Bbb R}^{N}$ it is possible to associate
to the Markov chain $\{x_{i_0},x_{i_1},\dots, x_{i_k}, \dots\}$ an
equivalent chain $\{X_{i_0},X_{i_1},\dots, X_{i_k}, \dots\}$ of
points specified through their coordinates in ${\Bbb R}^{N}$ but
still constrained to belong to the subset $V(X) = v$, that is, to
$\Sigma_{N\vb}$. 
Consequently, the invariant probability measure \cite{mcmc} of the Markov chain so constructed is the probability density (\ref{prob}).
Moreover, in the absence of critical points, for smooth functions
$F$ and smooth potentials $V$, the variation on each set $U_i$ of $F/\Vert\nabla V\Vert$ is  limited. Therefore, by keeping it finite  
the partition of unity can be refined as needed to make Lebesgue integration
convergent; hence equation (\ref{mcmcmc}) follows.$\quad\quad\square$

\begin{remark}
By introducing the following notation for the average of a generic measurable function $f:M^{N}\to\mathbb{R}$ over the
hypersurface $\LevelSetfunc{\bv}{\overline{V}_N}$ endowed with the measure $\D\mu^{N-1}_{\bv}$

\begin{equation}
\label{def:averageonSigmav}
\left\langle f \right\rangle_{\overline{v},\mu}=\dfrac{\displaystyle{\intSigmaV{\bv}{\overline{V}_N}\,f\D\mu_{\overline{v}}^{N-1}}}{\displaystyle{\intSigmaV{\bv}{\overline{V}_N}\,\D\mu_{\overline{v}}^{N-1}}}=
\dfrac{\displaystyle{\intSigmaV{\bv}{\overline{V}_N}\,f\D\mu_{\overline{v}}^{N-1}}}{\Omega_N(\overline{v})}\,\, ,
\end{equation}
 the quantities
\begin{equation}
\label{def: definition_statQuant}
\begin{split}
&\mathrm{Var}_{\overline{v},\mu}(f)=\mathrm{Cuml}^{(2)}_{\overline{v},\mu}(f)=\left\langle
f^2 \right\rangle_{\overline{v},\mu}-\left\langle f \right\rangle_{\overline{v},\mu}^2\\
&\mathrm{Cov}_{\overline{v},\mu}(f;g)=\left\langle f  g
\right\rangle_{\overline{v},\mu}-\left\langle
f\right\rangle_{\overline{v},\mu}\left\langle g \right\rangle_{\overline{v},\mu}\\
&\mathrm{Cuml}^{(3)}_{\overline{v},\mu}(f)=\left\langle f^3
\right\rangle_{\overline{v},\mu}-3\left\langle f
\right\rangle_{\overline{v},\mu}\left\langle f^2
\right\rangle_{\overline{v},\mu}+2 \left\langle f \right\rangle_{\overline{v},\mu}^3\\
&\mathrm{Cuml}^{(4)}_{\overline{v},\mu}(f)=\left\langle f^4 \right\rangle_{\overline{v},\mu}-4\left\langle f^3 \right\rangle_{\overline{v},\mu}\left\langle f
\right\rangle_{\overline{v},\mu}+12\left\langle f^2\right\rangle_{\overline{v},\mu}\left\langle f\right\rangle_{\overline{v},\mu}^2-3\left\langle f^2\right\rangle_{\overline{v},\mu}^2-6\left\langle f\right\rangle_{\overline{v},\mu}^4\\
\end{split}
\end{equation}
represent the variance, the correlation function, and the 3rd and 4th order 
cumulants on the hypersurface $\LevelSetfunc{\bv}{\overline{V}_N}$ with measure
$\D\mu^{N-1}_{\bv}$, respectively. 

With this notation, and substituting Eqs.\eqref{eq:Omega_Derivatives_Xi} in Eqs.\eqref{eq: microcanEntropy_Derivative}, the derivatives of the microcanonical
entropy at a non critical value $\bv$, and at fixed $N$ are worked out as averages of functions of $\overline{\zeta}_N= {\rm div} (\overline{\boldsymbol{\xi}}_N)$,  where the vector field $\overline{\boldsymbol{\xi}}_N$ generates the diffeomorphisms among the equipotential level sets, as follows

\begin{equation}
\begin{split}
&\dfrac{\D \overline{S}_N}{\D
\overline{v}}(\overline{v})=\dfrac{1}{N}\left\langle\overline{\zeta}_N\right\rangle_{\overline{v},\mu}\\
&\dfrac{\D^2 \overline{S}_N }{\D
\overline{v}^2}(\overline{v})=\dfrac{1}{N}\left[\mathrm{Var}_{\overline{v},\mu}(\overline{\zeta}_N)+\left\langle\mathcal{L}_{\overline{\boldsymbol{\xi}}_N}
(\overline{\zeta}_N)\right\rangle_{N\bv,\mu}\right]\\
&\dfrac{\D^3 \overline{S}_N }{\D\overline{v}^3}(\bv)=\dfrac{1}{N}\left[\mathrm{Cuml}^{(3)}_{\overline{v},\mu}(\overline{\zeta}_N)+3\mathrm{Cov}_{\overline{v},\mu}\left(\overline{\zeta}_N;\mathcal{L}_{\overline{\boldsymbol{\xi}}_{N}}(\overline{\zeta}_N)\right)+\left\langle\mathcal{L}_{\overline{\boldsymbol{\xi}}_N}^{(ii)}\left(\overline{\zeta}_N\right)\right\rangle_{\overline{v},\mu}\right]\\
&\dfrac{\D^4 \overline{S}_N }{\D\overline{v}^4}(\bv)=\dfrac{1}{N}\Biggr[\mathrm{Cuml}^{(4)}_{\overline{v},\mu}(\overline{\zeta}_N)+6\mathrm{Cov}_{\overline{v},\mu}\left(\overline{\zeta}_N^2;\mathcal{L}_{\overline{\boldsymbol{\boldsymbol{\xi}}}_N}(\overline{\zeta}_N)\right)+
3\mathrm{Var}_{\overline{v},\mu}\left(\mathcal{L}_{\overline{\boldsymbol{\boldsymbol{\xi}}}_N}(\overline{\zeta}_N)\right)+\\
&+4\mathrm{Cov}_{\overline{v},\mu}\left(\overline{\zeta}_N;\mathcal{L}_{\overline{\boldsymbol{\boldsymbol{\xi}}}_N}^{(ii)}(\overline{\zeta}_N)\right)
-12\left\langle\overline{\zeta}_N\right\rangle_{\overline{v},\mu}\mathrm{Cov}_{N\overline{v},\mu}\left(\overline{\zeta}_N;\mathcal{L}_{\overline{\boldsymbol{\boldsymbol{\xi}}}_N}(\overline{\zeta}_N)\right)+\left\langle\mathcal{L}_{\overline{\boldsymbol{\boldsymbol{\xi}}}_N}^{(iii)}\left(\overline{\zeta}_N\right)\right\rangle_{\overline{v},\mu}\Biggr]=\\
&=\dfrac{1}{N}\Biggr[\mathrm{Cuml}^{(4)}_{\overline{v},\mu}(\overline{\zeta}_N)+4\mathrm{Cov}_{\overline{v},\mu}\left(\overline{\zeta}_N;\mathcal{L}_{\overline{\boldsymbol{\boldsymbol{\xi}}}_N}^{(ii)}(\overline{\zeta}_N)\right)+3\mathrm{Var}_{\overline{v},\mu}\left(\mathcal{L}_{\overline{\boldsymbol{\boldsymbol{\xi}}}_N}(\overline{\zeta}_N)\right)+\\
&+6\left\langle\overline{\zeta}_N\right\rangle_{\overline{v},\mu}\left(\mathrm{Cov}_{\overline{v},\mu}\left(\Delta\overline{\zeta}_N;\mathcal{L}_{\overline{\boldsymbol{\boldsymbol{\xi}}}_N}(\overline{\zeta}_N)\right)\right)+\left\langle\mathcal{L}_{\overline{\boldsymbol{\boldsymbol{\xi}}}_N}^{(iii)}\left(\overline{\zeta}_N\right)\right\rangle_{\overline{v},\mu}
\Biggr]
\end{split}
\label{eq:microcanEntropy_Derivative_Xi}
\end{equation} 
where for sake of simplicity we have introduced the quantity
\begin{equation}
\Delta \overline{\zeta}_N=\dfrac{\overline{\zeta}_N^2}{\left\langle \overline{\zeta}_N \right\rangle_{\overline{v},\mu}}-2\overline{\zeta}_N \,\,\,.
\end{equation}
\end{remark}

Now the crucial step is to show that, under the hypothesis of diffeomorphicity that now includes \textit{asymptotic diffeomorphicity}, the function 
$\overline{\zeta}_N$ - considered along a MCMC spanning any given $\Sigma_v^{V_N}$ - is a Gaussian random process. This is achieved through an intermediate step to show that  the mean curvature $H$ - also considered along the same MCMC  - is a Gaussian random process.  For the sake of notation in what follows we shall omit the suffix $N$ of $V_N$.

\begin{lemma}[Mean curvature along a MCMC on a level set] The pointwise mean curvature of an $N$-dimensional manifold $\Sigma_{N\vb}^{V}$
\begin{equation}
H (\textit{\textbf{q}})= \frac{1}{N}\sum_{i=1}^N\kappa_i(\textit{\textbf{q}}) = - \frac{1}{N}\left[\frac{\Delta V}{\Vert\nabla V\Vert}
  - \frac{\partial^i V\partial^2_{ij} V \partial^j V}{\Vert\nabla V\Vert^3}\right]
\label{proprioH}
\end{equation}
computed along a Monte Carlo Markov Chain  $\{\textit{\textbf{q}}_k\}_{k\in{\mathbb N}}\in \Sigma_{N\vb}^{V}$ such that the stationary invariant density of the MCMC is the microcanonical configurational measure, 
where $\Sigma_{N\vb}^{V}$ is free of critical points of $V$, is a Gaussian random process.
\end{lemma}
{\bf Proof.} 
Along a MCMC, the principal curvatures $\kappa_i(\textit{\textbf{q}})$ behave as independent 
random variables with probability densities $u_i(\kappa_i)$ which we do not need to know explicitly. 
Statistical independence means that $\left \langle \kappa_i(\textit{\textbf{q}}) \kappa_j(\textit{\textbf{q}})\right \rangle^{\mu c}_{N,v} = \left \langle \kappa_i(\textit{\textbf{q}}) \right \rangle^{\mu c}_{N,v}\left \langle  \kappa_j(\textit{\textbf{q}})\right \rangle^{\mu c}_{N,v}$ and this can be understood as follows. 
Let $(M^n, g)$ be  an $n$-dimensional Riemannian manifold and $m$-codimensional submanifold of a Riemannian manifold $(\overline{M}^{m+n}, \overline g)$,  let $R$ and $\overline R$ denote the Riemann curvature tensors of $M^n$ and $\overline{M}^{m+n}$, respectively, and denote by $h(\cdot,  \cdot )$ the second fundamental form, then the Gauss equation reads
\begin{equation}
\overline {g}(\overline{ R} (X, Y) Z, W)) = g(R(X, Y) Z, W)) + \overline {g}(h(X,Z), h(Y, W)) - \overline {g}(h(X, W), h(Y, Z))
\label{gauss-eq}
\end{equation}
which, for sectional curvatures, obviously reads as
\begin{equation}
\overline {g}(\overline{ R} (X, Y) X, Y)) = g(R(X, Y) X, Y)) + \overline {g}(h(X,X), h(Y, Y)) - \overline {g}(h(X, Y), h(Y, X)) \ .
\label{gauss-eq1}
\end{equation}
Now, for any point $p\in M$ and basis $\{\textit{\textbf{e}}_1,\dots,\textit{\textbf{e}}_n\}$ of $T_pM$, it is possible to choose coordinates $(y^1,\dots,y^{n+1})$ in $\overline{M}$ such that the
tangent vectors $\textbf{Y}^1,\dots,\textbf{Y}^n$ coincide with 
$\{\textit{\textbf{e}}_1,\dots,\textit{\textbf{e}}_n\}$ and $\textit{\textbf{n}}=\textbf{Y}^{n+1}\in N_pM$ is orthogonal to $T_pM$.
Then $M$ is locally given as a graph manifold:  $y^1=x^1,\dots,y^n=x^n, y^{n+1} = f(\textit{\textbf{x}})$ so that the second fundamental form has the components \cite{secondaforma,thorpe}
\begin{equation}
h(\textit{\textbf{e}}_i,\textit{\textbf{e}}_j) = \frac{\partial^2f}{\partial x^i\partial x^j} \textit{\textbf{n}}
\label{gauss-eq3}
\end{equation}
where $\textit{\textbf{e}}_i = \partial /\partial x^i$.
Considering the potential level sets $\Sigma_{N\vb}^{V}$ as hypersurfaces of $\mathbb{R}^{N+1}$, identifying $f(x^1,\dots,x^N)$ with $V(q^1,\dots,q^N)$, taking $\textit{\textbf{n}}= \nabla V/\Vert\nabla V\Vert$, from Eqs.\eqref{gauss-eq},\eqref{gauss-eq1},\eqref{gauss-eq3} we obtain
\begin{equation}
K(\textit{\textbf{e}}_i,\textit{\textbf{e}}_j) = \kappa_i\kappa_j = - \left(\frac{\partial^2V}{\partial q_i^2}\right) \left( \frac{\partial^2V}{\partial q_j^2}\right)\langle \textit{\textbf{n}},\textit{\textbf{n}}\rangle + \left(\frac{\partial^2V}{\partial q_i\partial q_j} \right)^2 \langle \textit{\textbf{n}},\textit{\textbf{n}}\rangle
\label{sectional}
\end{equation}
hence
\begin{equation}
\left \langle \kappa_i(\textit{\textbf{q}}) \kappa_j(\textit{\textbf{q}})\right \rangle^{\mu c}_{N,v} = \left \langle \frac{1}{\Vert\nabla V\Vert}\left[  \left(\frac{\partial^2V}{\partial q_i\partial q_j} \right)^2 - 
\left(\frac{\partial^2V}{\partial q_i^2}\right) \left( \frac{\partial^2V}{\partial q_j^2}\right) \right] \right \rangle^{\mu c}_{N,v} \ .
\label{sectional1}
\end{equation}
For short-range interactions with coordination number $n_0$, meaning that - with a suitable labelling of the variables - $q_i$ and $q_{j}$ do not interact if $\vert i - j\vert > n_0$, the entries of the Hessian of $V$ vanish if $\vert i - j\vert > n_0$. Thus, locally, for $n > n_0$, we have
\begin{equation}
\left \langle \kappa_i(\textit{\textbf{q}}) \kappa_{j=i+n}(\textit{\textbf{q}})\right \rangle^{\mu c}_{N,v} = \left \langle - \left[ \frac{1}{\Vert\nabla V\Vert^{1/2}}\left(\frac{\partial^2V}{\partial q_i^2}\right)  \right]  \left[ \frac{1}{\Vert\nabla V\Vert^{1/2}}\left( \frac{\partial^2V}{\partial q_j^2}\right) \right] \right \rangle^{\mu c}_{N,v} \ .
\label{sectional2}
\end{equation}
with evident notation, we can write 
\begin{equation}
\left \langle \kappa_i(\textit{\textbf{q}}) \kappa_{j=i+n}(\textit{\textbf{q}})\right \rangle^{\mu c}_{N,v} = \left \langle \left[ \left \langle \kappa_i(\textit{\textbf{q}}) \right \rangle^{\mu c}_{N,v} + \delta \kappa_i(\textit{\textbf{q}}) \right] \left[\left \langle \kappa_{j=i+n}(\textit{\textbf{q}})\right \rangle^{\mu c}_{N,v} + \delta \kappa_{j=i+n}(\textit{\textbf{q}}) \right]
\right \rangle^{\mu c}_{N,v} \ .
\label{sectional3}
\end{equation}
As we shall see in the following, $1/\Vert\nabla V\Vert^{1/2}$ tends to a constant value at increasing $N$, and along a MCMC sampling a potential level set with its configurational microcanonical measure, the fluctuations of $\partial^2V/\partial q_i^2$ and $\partial^2V/\partial q_j^2$ are independent and average to zero. In conclusion, under the condition $\vert i - j\vert > n_0$ we have
\[
\left \langle \kappa_i(\textit{\textbf{q}}) \kappa_j(\textit{\textbf{q}})\right \rangle^{\mu c}_{N,v} = \left \langle \kappa_i(\textit{\textbf{q}}) \right \rangle^{\mu c}_{N,v}\left \langle  \kappa_j(\textit{\textbf{q}})\right \rangle^{\mu c}_{N,v} \ .
\] 
Having shown that the principal curvatures $\kappa_i(\textit{\textbf{q}})$ are everywhere uniformly bounded on any $\Sigma_{N\vb}^{V}$ belonging to any cylindrical subset of the family  $\left\{\Gamma^N_{{I}_0}\right\}_{N\in\mathbb{N}}$, the consequence is that the momenta of the distributions $u_i(\kappa_i)$ are finite and uniformly bounded too. Hence, the basic conditions  are fulfilled to apply the Central Limit Theorem (CLT) formulated by Khinchin \cite{khinchin} for sum functions  of independent random variables, arbitrarily distributed, with bounded momenta up to the fifth order. Hence,  along the MCMC $\{\textit{\textbf{q}}_k\}_{k\in{\mathbb N}}\in \Sigma_{N\vb}^{V}$ the invariant measure of which is the configurational microcanonical one, the values of the mean curvature $H (\textit{\textbf{q}}_k)$ behave as Gaussian-distributed random variables. Notice that a finite range dependence is a weak dependence that does not prevent the CLT to apply \cite{clt-weak}.

\begin{lemma}[$\zeta_N(\textit{\textbf{q}})$ along the MCMC on regular level sets] The quantity
\begin{equation}\label{quasiH}
{\zeta}_N(\textit{\textbf{q}}) = \frac{\Delta V}{\Vert\nabla V\Vert^2}
  -2 \frac{\partial^i V\partial^2_{ij} V \partial^j V}{\Vert\nabla V\Vert^4} 
\end{equation}
as well as $\overline{\zeta}_N(\textit{\textbf{q}})$, computed along a Monte Carlo Markov Chain $\{\textit{\textbf{q}}_k\}_{k\in{\mathbb N}}\in \Sigma_{N\vb}^{V}$ the invariant measure of which is the configurational microcanonical one, is a Gaussian random process. 
\end{lemma}
{\bf Proof.} After the preceding Lemma it follows that the two quantities ${\Delta V}/{\Vert\nabla V\Vert}$ and ${\partial^i V\partial^2_{ij} V \partial^j V}/{\Vert\nabla V\Vert^3}$ -- computed along Monte Carlo Markov Chain $\{\textit{\textbf{q}}_k\}_{k\in{\mathbb N}}\in \Sigma_{N\vb}^{V}$ the invariant measure of which is the configurational microcanonical one --- are gaussian random processes because their sum is a gaussian random process and the sum of gaussian random processes is gaussian. Now, if the quantity $\frac{\Delta V}{\Vert\nabla V\Vert}= \sum_i \partial_{ii}^2V/{\Vert\nabla V\Vert}$ is asymptotically gaussian, then  the terms  $\partial_{ii}^2V/{\Vert\nabla V\Vert}$ have to be i.i.d. random variables as well the terms $\partial_{ii}^2V$ because all of them are divided by the same number ${\Vert\nabla V\Vert}$ at each point of the MCMC, by the same token $\partial_{ii}^2V/{\Vert\nabla V\Vert^2}$ have to be i.i.d. random variables because now all the terms $\partial_{ii}^2V$ are divided by the same number ${\Vert\nabla V\Vert^2}$ at each point of the MCMC.
The same argument applies to ${\partial^i V\partial^2_{ij} V \partial^j V}/{\Vert\nabla V\Vert^3}$ so that in the end both 
 ${\Delta V}/{\Vert\nabla V\Vert^2}$ and ${\partial^i V\partial^2_{ij} V \partial^j V}/{\Vert\nabla V\Vert^4}$ are gaussian distributed, and, consequently, 
 $\overline{\zeta}_N(\textit{\textbf{q}})$ in Eq.\eqref{quasiH} is a gaussian random variable along a MCMC the invariant measure of which is the configurational microcanonical one.
 
 \begin{remark}. 
 Let us emphasize that the quantity $\overline{\zeta}_N(\textit{\textbf{q}})$ is a random variable along all the MCMC $\{\textit{\textbf{q}}_k\}_{k\in{\mathbb N}}\in \Sigma_{N_k\vb}^{V_{N_k}}$, with vanishing deviations from a gaussian distribution at increasing $N$,  under the hypothesis of asymptotic diffeomorphicity because the principal curvatures $\kappa_i(\textit{\textbf{q}})$ are uniformly bounded with $N$ from above on any manifold, a crucial condition for the validity of Lemma 2.
 \end{remark}
 
 \begin{remark}. In the hypotheses of the Main Theorem, $V$ contains only short-range interactions and
its functional form does not change with $N$. In other words, we are tackling physically
homogeneous systems, which, at any $N$, can be considered as the union of smaller and identical subsystems. 
If a system is partitioned into a number $k$ of sufficiently large subsystems, the larger $N$ the more accurate
the factorization of its configuration space. Therefore, the averages of functions of interacting
variables belonging to a given block depend neither on the subsystems where they are
computed (the potential functions are the same on each block after suitable relabelling of the variables)
nor on the total number $N$ of degrees of freedom.
 \begin{itemize}
 \item[ ] {a)} Since the
potential $V$ is assumed smooth and bounded below, one has
\bena
\langle \mid \Delta V \mid \rangle^{\mu c}_{N,v}= \left \langle
\left \vert \sum_{i=1}^N \partial_{ii}^2 V \right \vert \right
\rangle^{\mu c}_{N,v}\ \leq \sum_{i=1}^N \langle \mid
\partial_{ii}^2 V \mid \rangle^{\mu c}_{N,v} \ \leq
N~\max_{i=1,\dots,N} \left \langle \left ( \mid \partial_{ii}^2 V
\mid \right ) \right \rangle^{\mu c}_{N,v}\ . \nonumber
\eena
At large $N$ (when the
fluctuations of the averages are vanishingly small)
$\max_{i=1,\dots,N}\langle \mid \partial_{ii}^2 V \mid \rangle^{\mu
c}_{N,v}$ does not depend on $N$, and  the same holds for $\left
\langle \mid \partial^i V\partial^2_{ij} V \partial^j V
\mid\right\rangle^{\mu c}_{N,v}$ and $\max_{i,j =1,\dots,N} \left
\langle \mid \partial^i V\partial^2_{ij} V \partial^j
V \mid \right \rangle^{\mu c}_{N,v}$.\\
Hence we set 
\[
m_1=\max_{i=1,\dots,N} \langle \mid \partial_{ii}^2 V \mid
\rangle^{\mu c}_{N,v}
\] 
\begin{equation}\label{m1m2}
m_2=\max_{i,j=1,\dots,N} \left \langle
\mid \partial^i V\partial^2_{ij} V
\partial^j V \mid \right \rangle^{\mu c}_{N,v} \ .
\end{equation}

\item[ ] {b)} Moreover, the absence of critical points of $V$, implied by the hypothesis of diffeomorphicity of the equipotential hypersurfaces, means that $\Vert\nabla V\Vert^2\geq C>0$. Hence the terms $\langle \ngV^{2n} \rangle^{\mu c}_{N,v}$ for $n=1,\dots , 8$ we have 
\bena
\langle \ngV^2 \rangle^{\mu
c}_{N,v}&=& \left\langle\sum_{i=1}^N ( \partial_i V )^2
\right\rangle^{\mu c}_{N,v} \ = \sum_{i=1}^N \left\langle (
\partial_i V )^2\right\rangle^{\mu c}_{N,v} \geq N~\min_{i=1,\dots,N}
\left \langle \left( \partial_i V \right)^2 \right \rangle^{\mu
c}_{N,v}\ , \nonumber\\
 \langle \ngV^4 \rangle^{\mu
c}_{N,v}&=& \left\langle \left[\sum_{i=1}^N ( \partial_i V
)^2\right]^2 \right\rangle^{\mu c}_{N,v} \ = \sum_{i,j=1}^N
\left\langle ( \partial_i V )^2( \partial_j V
)^2\right\rangle^{\mu c}_{N,v}
\nonumber\\
&\geq & N^2~\min_{i,j=1,..,N} \left \langle \left( \partial_i V
\right)^2 \left( \partial_j V \right)^2 \right \rangle^{\mu
c}_{N,v}\ ,
\nonumber
\eena
which can be iterated up to $\langle \ngV^{16} \rangle^{\mu c}_{N,v}$
By setting
\[
c_1=\min_{i=1,\dots,N}\left
\langle \left( \partial_i V \right)^2 \right \rangle^{\mu
c}_{N,v}
\]
\[
c_2=\min_{i,j=1,..,N} \left \langle \left(
\partial_i V \right)^2 \left( \partial_j V \right)^2 \right
\rangle^{\mu c}_{N,v}
\]
\[
............................
\]
\begin{equation}\label{c1c8}
c_8=\min_{{i_1},\dots,{i_8}=1,..,N} \left \langle \left(
\partial_{i_1} V \right)^2 \dots\left( \partial_{i_8} V \right)^2 \right
\rangle^{\mu c}_{N,v}
\end{equation}
\item[ ] {c)} By the same token put forward at the beginning of this Remark, we can define the following
quantities independent of $N$
\begin{eqnarray}\label{m3m7}
m_3&=&\max_{i,j,k,l=1,N}\left\langle (\de_i V \de^2_{ij}V \de_j V)
(\de_k V \de^2_{kl}V \de_l V) \right \rangle_{N,v}^{\mu c}\ ,
\nonumber\\
m_4&=&\max_{i,j,k=1,N}\left\langle \de_i V \de^2_{ij}V \de^2_{jk}V
\de_k V
 \right \rangle_{N,v}^{\mu c}\ ,
\nonumber\\
m_5&=&\max_{i,j,k=1,N}\left\langle (\de_i V \de^2_{ij}V \de_j V)
(\de^2_{kk}V )  \right \rangle_{N,v}^{\mu c}\ ,
\nonumber\\
m_6&=&\max_{i,j=1,N}\left\langle \de_i V \de^3_{ijj}V  \right
\rangle_{N,v}^{\mu c}\ ,
\nonumber\\
m_7&=&\max_{i,j,k=1,N}\left\langle (\de_i V \de_jV \de_k V)
\de^3_{ijk}V   \right \rangle_{N,v}^{\mu c}\ ,\nonumber\\
m_8&=&\max_{i,j=1,\dots,N} \left \langle
\mid \partial^2_{ij} V
\partial^j V \mid \right \rangle^{\mu c}_{N,v} \ .
\end{eqnarray}
\end{itemize}
 \end{remark}

 \begin{lemma}[Upper bound of the first derivative of the entropy] 
 After Lemmas 2 to 5 and Remarks 2 to 5 it is
\[
\sup_{N,\vb\in {I_{\vb}}} \left | \frac{\de S_N}{\de
{\vb}}({\vb}) \right | < \infty
 \]
 \end{lemma}
 \textbf{Proof.} {\rm 
This first derivative of the entropy is equal to the inverse of the configurational temperature, thus it is necessarily uniformly bounded with $N$. In fact, the property of $\overline{\zeta}_N(\textit{\textbf{q}})$ - and of   ${\zeta}_N(\textit{\textbf{q}})$ - of being a Gaussian distributed random variable along any MCMC defined above entails the following uniform bound }
\begin{equation}
\lim_{N\rightarrow+\infty}\left\langle \zeta_N \right\rangle_{N\bv,\mu}=\lim_{N\rightarrow+\infty}N^{-1}\left\langle \overline{\zeta}_N \right\rangle_{N\bv,\mu}\in\mathbb{R}\ .  \hskip 1truecm \square
\end{equation}
 \begin{lemma}[Upper bound of the second derivative of the entropy] 
 After Lemmas 2 to 5 and Remarks 2 to 5 it is
\begin{equation}
\sup_{N,\vb\in {I_{\vb}}} \left\vert \frac{\de^2 S_N}{\de {\vb}^2}({\vb}) \right\vert < \infty
\end{equation}
 \end{lemma}
 \textbf{Proof.} {\rm 
Since $\overline{\zeta}_N(\textit{\textbf{q}})$ is a gaussian random process,  the quantity $\mathrm{Var}_{\overline{v},\mu}(\overline{\zeta}_N)/N$ is 
uniformly bounded with $N$. Then the $N$-dependence of the average  $\left\langle\mathcal{L}_{\overline{\boldsymbol{\xi}}_N}(\overline{\zeta}_N)\right\rangle_{N\bv,\mu}$ follows from the explicit expression of the quantity $\mathcal{L}_{\overline{\boldsymbol{\xi}}_N}(\overline{\zeta}_N)$  given by Eq.\eqref{derivLie} in Appendix B (by adapting it to quantities marked with an overline). Considering that the number of non-vanishing entries of the Hessian of the potential is ${\cal O}(n_pN)$, where $n_p$ is the number of nearest-neighbors in condensed matter systems and the average number of neighbors of a particle in a fluid system, using the above defined $N$-independent quantities in Eqs.\eqref{m1m2},\eqref{c1c8},\eqref{m3m7} a simple estimation term by term gives
\begin{equation}
\label{derivLiebound}
\begin{split}
\left\langle \mathcal{L}_{\overline{\boldsymbol{\xi}}_N}(\overline{\zeta}_N)\right\rangle_{N\bv,\mu}&\leq \left\langle\left\vert\dfrac{\nabla \overline{V}\cdot\nabla(\Delta \overline{V})}{\|\nabla \overline{V}\|^4}\right\vert\right\rangle_{N\bv,\mu} +\left\langle\left\vert 8\dfrac{(\nabla \overline{V} \cdot\Hess \overline{V}\nabla \overline{V})^2}{\|\nabla \overline{V}\|^8}\right\vert\right\rangle_{N\bv,\mu} + \\
&+ \left\langle\left\vert2\dfrac{(\nabla \overline{V} \cdot\Hess(\overline{V})\nabla \overline{V})\Delta \overline{V}+2\|\Hess \overline{V}\nabla \overline{V}\|^2+\mathrm{D}^3\overline{V}(\nabla \overline{V},\nabla \overline{{V}},\nabla \overline{V})}{\|\nabla \overline{V}\|^6}   \right\vert\right\rangle_{N\bv,\mu}\\
&\leq N\dfrac{m_6}{c_2} +  8\dfrac{m_2^2 n_p^2}{c_4} + 2N\dfrac{m_5 n_p +2m_8^2 n_p^2 +m_7 n_p}{c_3} 
\end{split}
\end{equation}
that is, the upper bound of  $\left\langle\mathcal{L}_{\overline{\boldsymbol{\xi}}_N}(\overline{\zeta}_N)\right\rangle_{N\bv,\mu}/N$ of this quantity remains uniformly bounded in the $N\to\infty$ limit. }
$\square$

\begin{lemma} [Upper bound of the third derivative of the entropy] {\rm 
 After Lemmas 2 to 5 and Remarks 2 to 5 it is}
\[
\sup_{N,\vb\in {I_{\vb}}} \left | \frac{\de^3 S_N}{\de
{\vb}^3}({\vb}) \right | < \infty
\]
\end{lemma}
 \textbf{Proof.} {\rm  
 Since $\overline{\zeta}_N(\textit{\textbf{q}})$ is a gaussian random process, we have the following uniform bound
\[
 \lim_{N\rightarrow+\infty}N^{2}\mathrm{Cumul}^{(3)}_{N\bv,\mu}\zeta_N=\lim_{N\rightarrow+\infty}N^{-1}\mathrm{Cumul}^{(3)}_{N\bv,\mu} \overline{\zeta}_N=0 \ ,
\]
 and by considering the explicit expression of $\mathcal{L}^{(ii)}_{{\boldsymbol{\xi}}_N}({\zeta}_N)$  given by Eq.\eqref{deriv2Lie} in 
 Appendix B, a tedious but trivial counting of the $N$-dependence  term by term of $\mathcal{L}^{(ii)}_{\overline{\boldsymbol{\xi}}_N}(\overline{\zeta}_N)$  - as in the previous case - shows that $\left\langle\mathcal{L}^{(ii)}_{\overline{\boldsymbol{\xi}}_N}(\overline{\zeta}_N)\right\rangle_{N\bv,\mu}$ turns out of order ${\cal O}(n_p^3N)$ and thus divided by $N$ remains uniformly bounded in the $N\to\infty$ limit.  Then, according to the definition \eqref{def: definition_statQuant} we have
\[
\mathrm{Cov}_{\overline{v},\mu}\left(\overline{\zeta}_N;\mathcal{L}_{\overline{\boldsymbol{\xi}}_{N}}(\overline{\zeta}_N)\right) = 
\left\langle \overline{\zeta}_N\  \mathcal{L}_{\overline{\boldsymbol{\xi}}_{N}}(\overline{\zeta}_N) \right\rangle_{\overline{v},\mu}
-  \left\langle \overline{\zeta}_N  \right\rangle _{\overline{v},\mu}  
\left\langle \mathcal{L}_{\overline{\boldsymbol{\xi}}_{N}}(\overline{\zeta}_N)  \right\rangle_{\overline{v},\mu }
\]
where the quantities $\overline{\zeta}_N $  and $\mathcal{L}_{\overline{\boldsymbol{\xi}}_{N}}(\overline{\zeta}_N)$ are randomly varying along the MCMC whose probability measure is the configurational microcanonical measure, and the random variations of $\overline{\zeta}_N $ and of its directional (Lie) derivative in a random direction ${\boldsymbol{\xi}}_{N}$ can be considered \textit{bona fide} statistically uncorrelated, thus their covariance vanishes.} $\square$
\begin{lemma}[Upper bound of the fourth derivative of the entropy] 
 After Lemmas 2 to 5 and Remarks 2 to 5 it is
\[
\sup_{N,\vb\in {I_{\vb}}} \left | \frac{\de^4 S_N}{\de
{\vb}^4}({\vb}) \right | < \infty
\]
 \end{lemma}
 \textbf{Proof.} {\rm  
Since $\overline{\zeta}_N(\textit{\textbf{q}})$ is a gaussian random process, we have the following uniform bound
\[
\lim_{N\rightarrow+\infty}N^{3}\mathrm{Cumul}^{(4)}_{N\bv,\mu}\zeta_N =\lim_{N\rightarrow+\infty}N^{-1}\mathrm{Cumul}^{(4)}_{N\bv,\mu}\overline{\zeta}_N
=0\ .
\]
Then, by considering the expression of  $\mathcal{L}^{(iii)}_{{\boldsymbol{\xi}}_N}({\zeta}_N)$  given by Eq.\eqref{deriv3Lie} in 
 Appendix B, a very tedious but trivial counting of the $N$-dependence  term by term of  $\mathcal{L}^{(iii)}_{\overline{\boldsymbol{\xi}}_N}(\overline{\zeta}_N)$ - as in the previous case - shows that $\left\langle\mathcal{L}^{(iii)}_{\overline{\boldsymbol{\xi}}_N}(\overline{\zeta}_N)\right\rangle_{N\bv,\mu}$ turns out of order ${\cal O}(n_p^4N)$ and thus divided by $N$ remains uniformly bounded in the $N\to\infty$ limit.

Now, the term $N^{-1}\mathrm{Var}_{\overline{v},\mu}\left(\mathcal{L}_{\overline{\boldsymbol{\boldsymbol{\xi}}}_N}(\overline{\zeta}_N)\right)$ is also uniformly bounded, in fact along the MCMC spanning a given
equipotential surface the terms $\overline{\xi}_i\partial_i {\overline{\zeta}}$ are random uncorrelated variables each bringing a factor $N$ because ${\overline{\zeta}}\sim {\cal O}(N)$, thus
\begin{eqnarray}
\mathrm{Var}_{\overline{v},\mu}\left(\mathcal{L}_{\overline{\boldsymbol{\boldsymbol{\xi}}}_N}(\overline{\zeta}_N)\right)&=&\mathrm{Var}_{\overline{v},\mu}\left(
\sum_{i=1}^N{\overline{\xi}}_i\partial_i {\overline{\zeta}}\right)= \mathrm{Var}_{\overline{v},\mu}\left(\frac{1}{N}\sum_{i=1}^N{N \overline{\xi}}_i\partial_i \overline{\zeta} \right) = \frac{1}{N^2}\sum_{i=1}^N\mathrm{Var}_{\overline{v},\mu}\left(N \overline{\xi}_i\partial_i {\overline{\zeta}}\right)\nonumber\\
&\le&\frac{1}{N^2}N \sigma_m^2N^2 = N \sigma_m^2
\end{eqnarray}
where $\sigma_m^2$ is the largest value of all the standard deviations of the terms
$\overline{\xi}_i\partial_i \overline{\zeta}$ along the MCMC. 

For what concerns the two remaining terms in the fourth derivative of the entropy in Eq.\eqref{eq:microcanEntropy_Derivative_Xi} we have 
\begin{equation}\label{cov1}
\mathrm{Cov}_{\overline{v},\mu}\left(\overline{\zeta}_N;\mathcal{L}_{\overline{\boldsymbol{\xi}}_{N}}^{(ii)}(\overline{\zeta}_N)\right) = 
\left\langle \overline{\zeta}_N\  \mathcal{L}_{\overline{\boldsymbol{\xi}}_{N}}^{(ii)}(\overline{\zeta}_N) \right\rangle_{\overline{v},\mu}
-  \left\langle \overline{\zeta}_N  \right\rangle _{\overline{v},\mu}  
\left\langle \mathcal{L}_{\overline{\boldsymbol{\xi}}_{N}}^{(ii)}(\overline{\zeta}_N)  \right\rangle_{\overline{v},\mu }
\end{equation}
that vanishes when computed as microcanonical averages through “time” averages along a MCMC, in fact, we take advantage of the resulting complete decorrelation between the random values taken by $\overline{\zeta}_N$ and the random values of its second order Lie derivative taken in a random direction 
${\boldsymbol{\xi}}_N$.
\begin{equation}\label{cov2}
\mathrm{Cov}_{\overline{v},\mu}\left(\Delta\overline{\zeta}_N;\mathcal{L}_{\overline{{\boldsymbol{\xi}}}_N}(\overline{\zeta}_N)\right) =
\left\langle\Delta\overline{\zeta}_N\ \mathcal{L}_{\overline{\boldsymbol{\boldsymbol{\xi}}}_N}(\overline{\zeta}_N)\right\rangle - \left\langle\Delta\overline{\zeta}_N\right\rangle \left\langle\mathcal{L}_{\overline{\boldsymbol{\boldsymbol{\xi}}}_N}(\overline{\zeta}_N)\right\rangle
\end{equation}
the same argument applies to the quantities  $\Delta\overline{\zeta}_N$ and $\mathcal{L}_{\overline{{\boldsymbol{\xi}}}_N}(\overline{\zeta}_N)$ that are uncorrelated random variables along a MCMC, thus their covariance vanishes. $\quad\square$ }

\section{Fixing the problem raised by the lattice $\phi^4$ model} 

A few years ago, an argument was raised \cite{kastner} against the topological theory of phase transitions on the basis of the observation that the second order phase transition of the $2D$ lattice $\phi^4$-model occurs at a critical value $v_c$ of the potential energy density which belongs to a broad interval of $v$-values void of critical points of the potential function. In other words, for any finite $N$ the $\{ \Sigma_{v<v_c}^{V_N}\}_{v \in{\mathbb R}}$ are diffeomorphic to the  $\{ \Sigma_{v>v_c}^{V_N}\}_{v \in{\mathbb R}}$ so that no topological change seems to correspond to the phase transition. This is a counterexample to the theorem in Refs. \cite{prl1,NPB1}.  A first reply was given in \cite{vaff} where, in spite of the absence of critical points of the potential in
correspondence of the transition energy, a strong evidence has been given to relate the phase transition of this model with a change of topology of both the energy and potential level sets. But in this case the topology changes are asymptotic ($N\to\infty$). 

Let us see how the Main Theorem proved in the present work definitely fixes the problem, so that the $2D$ lattice $\phi^4$-model is no longer a counterexample to the topological necessity theorem.

The model of interest, considered in Ref.\cite{CCP},  is defined by the Hamiltonian 
\begin{equation}
{\cal H}_{N} ( p, q )= \sum_{{\textit{\textbf{i}} }}   \frac{p_{{\textit{\textbf{i}}}}^2}{2}  + V_{N}(q)
\label{Hphi_2}
\end{equation} 
where the potential function $V(q)$ is
\begin{equation}
V(q)=\sum_{{\textit{\textbf{i}}}\in{\mathbb Z}^D}\left( - \frac{m^2}{2} q_{{\textit{\textbf{i}}}}^2 +
\frac{\lambda}{4!} q_{{\textit{\textbf{i}}}}^4 \right) + \sum_{\langle {{\textit{\textbf{ik}}}\rangle\in{\mathbb Z}^D}} \frac{1}{2}J (q_{{\textit{\textbf{i}}}}-q_{{\textit{\textbf{k}}}})^2\ ,
\label{potfi4}
\end{equation}
with $\langle {\textit{\textbf{ik}}}\rangle$ standing for nearest-neighbor sites on a $D$ dimensional lattice. This
system has a discrete ${\mathbb Z}_2$-symmetry and short-range
interactions; therefore, according to the Mermin--Wagner theorem,
in $D=1$ there is no phase transition whereas in $D=2$ there is a a second order 
symmetry-breaking transition,  with nonzero critical temperature, of the same
universality class of the 2D Ising model. \\
In this Section we present the results of Monte Carlo numerical simulations on equipotential level
set of this model on a 2D-lattice with periodic boundary conditions and the following parameters: $J=1$, $m^2 = 2$, and $\lambda = 0.6$. For these values  of the parameters, the $2D$ system undergoes the symmetry-breaking phase transition at the critical potential energy density value is $v_c=\langle V\rangle_c/N\simeq 2.2$. 
This study has been performed in order to identify which terms composing the derivatives of the specific configurational microcanonical entropy  with respect to the specific potential energy is not uniformly bounded in $N$, as is expected after the present Main Theorem.\\
The simulations have been performed for systems with different number of degrees of freedom: $N=10\times 10=100$, $N=20 \times 20=400$, $N=30\times 30=900$, $N=40\times 40=1600$, $N=50\times 50=2500$ and $N=70\times 70=4900$. The computations were performed with vanishing magnetization as initial condition, for $2\times 10^{7}$ steps, a number sufficient to guarantee the convergence of the reported quantities.\\

\begin{figure}[h!]
 \includegraphics[scale=0.2,keepaspectratio=true]{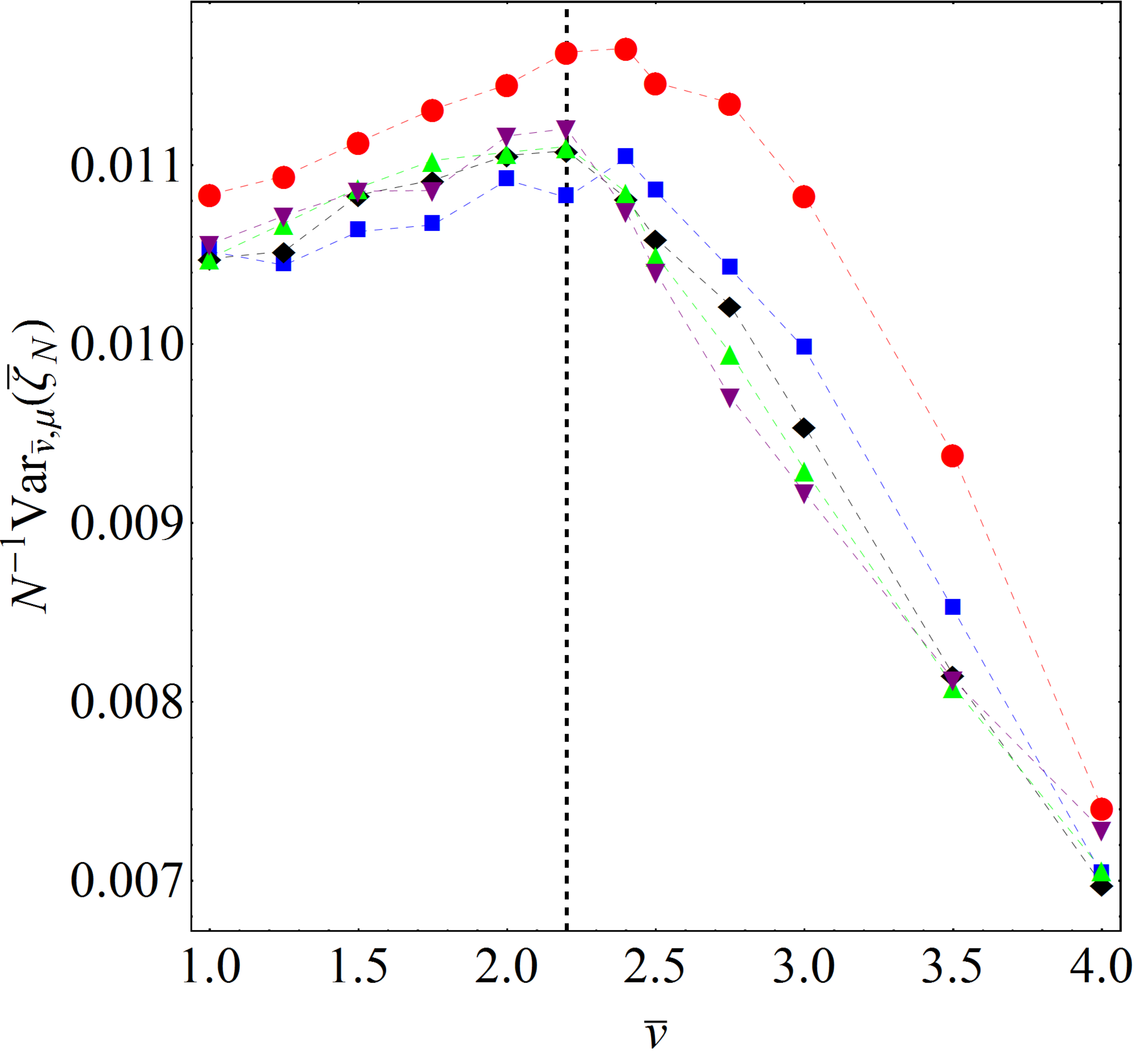}\hskip 1truecm
\includegraphics[scale=0.215,keepaspectratio=true]{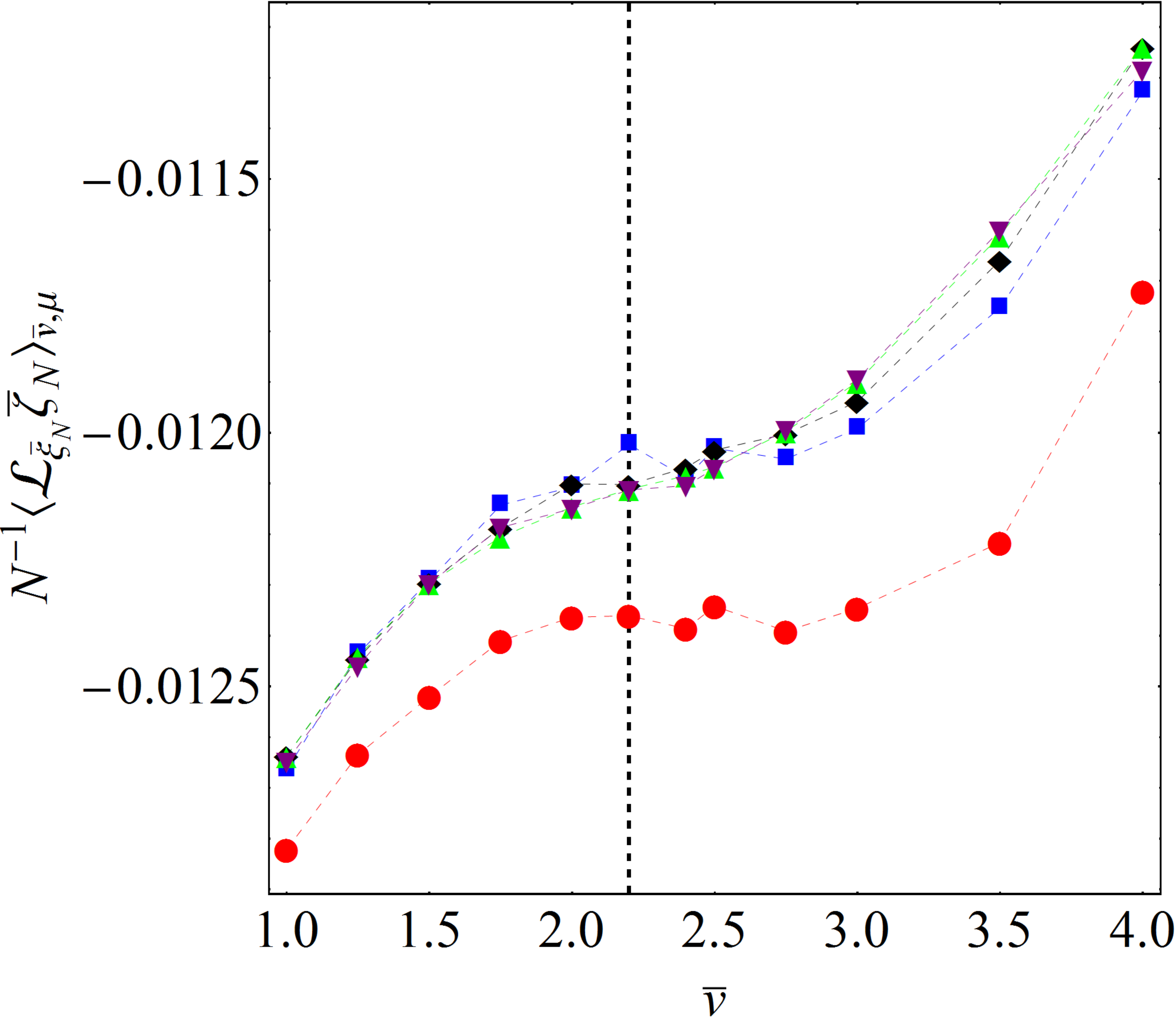}
\caption[10pt]{Quantites entering Eq.\protect\eqref{eq:microcanEntropy_Derivative_Xi} for lattices with different $N$. In particular, $N=100$ (red full circles), $N=400$ (blue squares), $N=900$ (black diamonds), $N=1600$ (green triangles), $N=2500$ (purple reversed triangles). }
\label{VarLieXiZeta}
\end{figure}  
\begin{figure}[h!]
\includegraphics[scale=0.2,keepaspectratio=true]{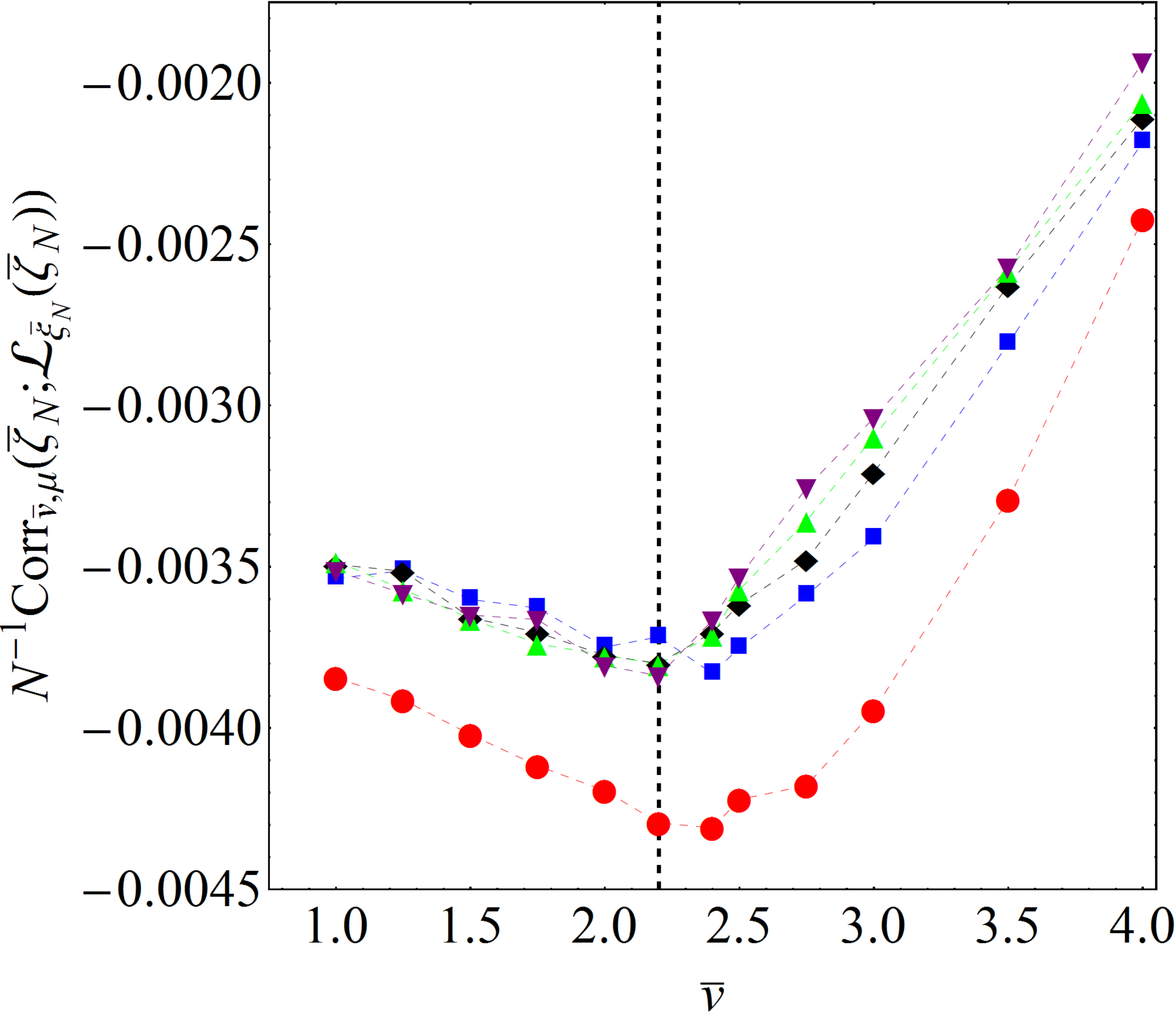}\hskip 1truecm
\includegraphics[scale=0.2,keepaspectratio=true]{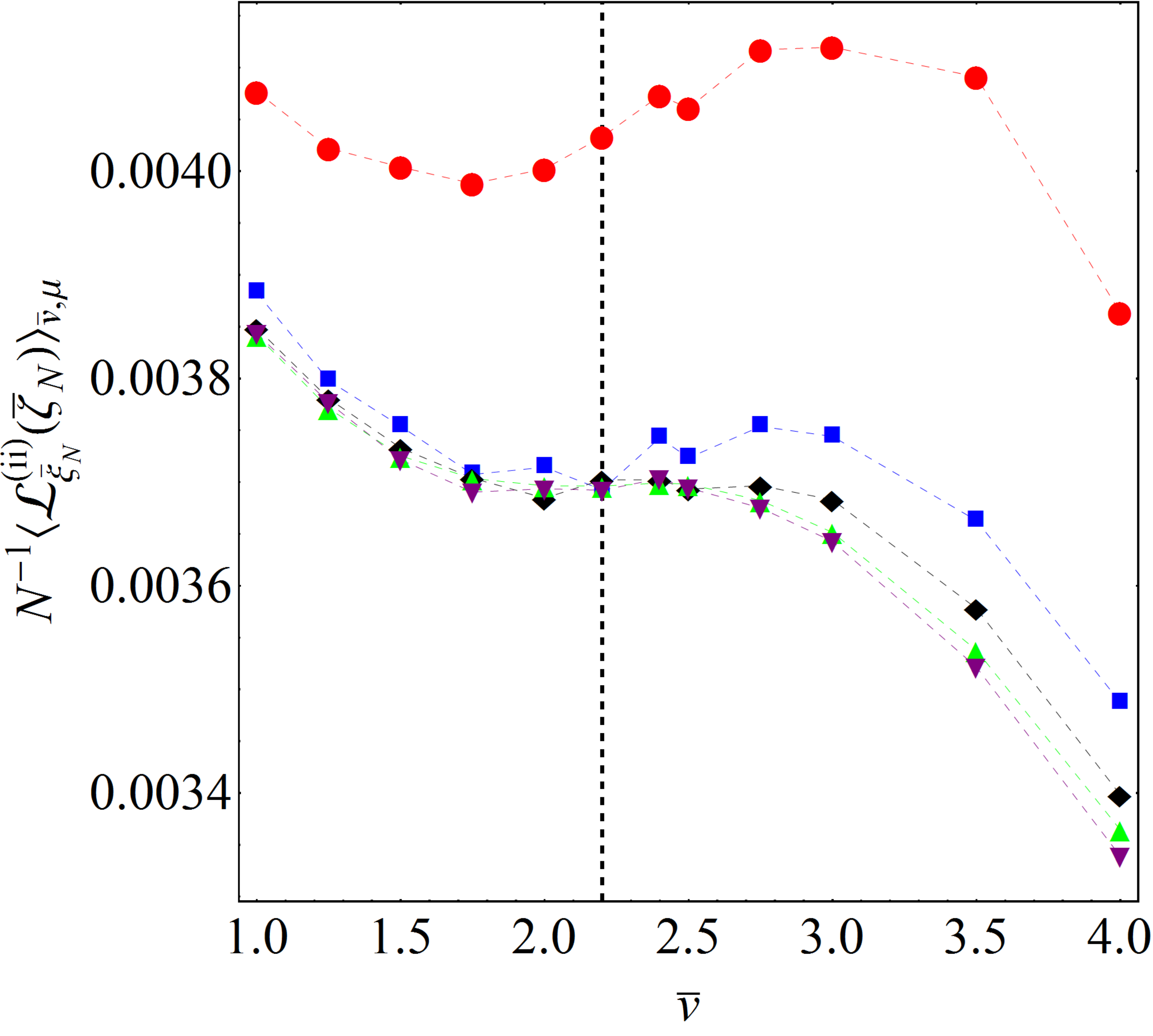}
\caption[10pt]{Quantites entering Eq.\protect\eqref{eq:microcanEntropy_Derivative_Xi} for lattices with different $N$. In particular, $N=100$ (red full circles),$N=400$ (blue squares), $N=900$ (black diamonds), $N=1600$ (green triangles), $N=2500$ (purple reversed triangles). }
\label{CorrZeta_LieXiZeta}
\end{figure}  
 
\begin{figure}[h!]
 \includegraphics[scale=0.2,keepaspectratio=true]{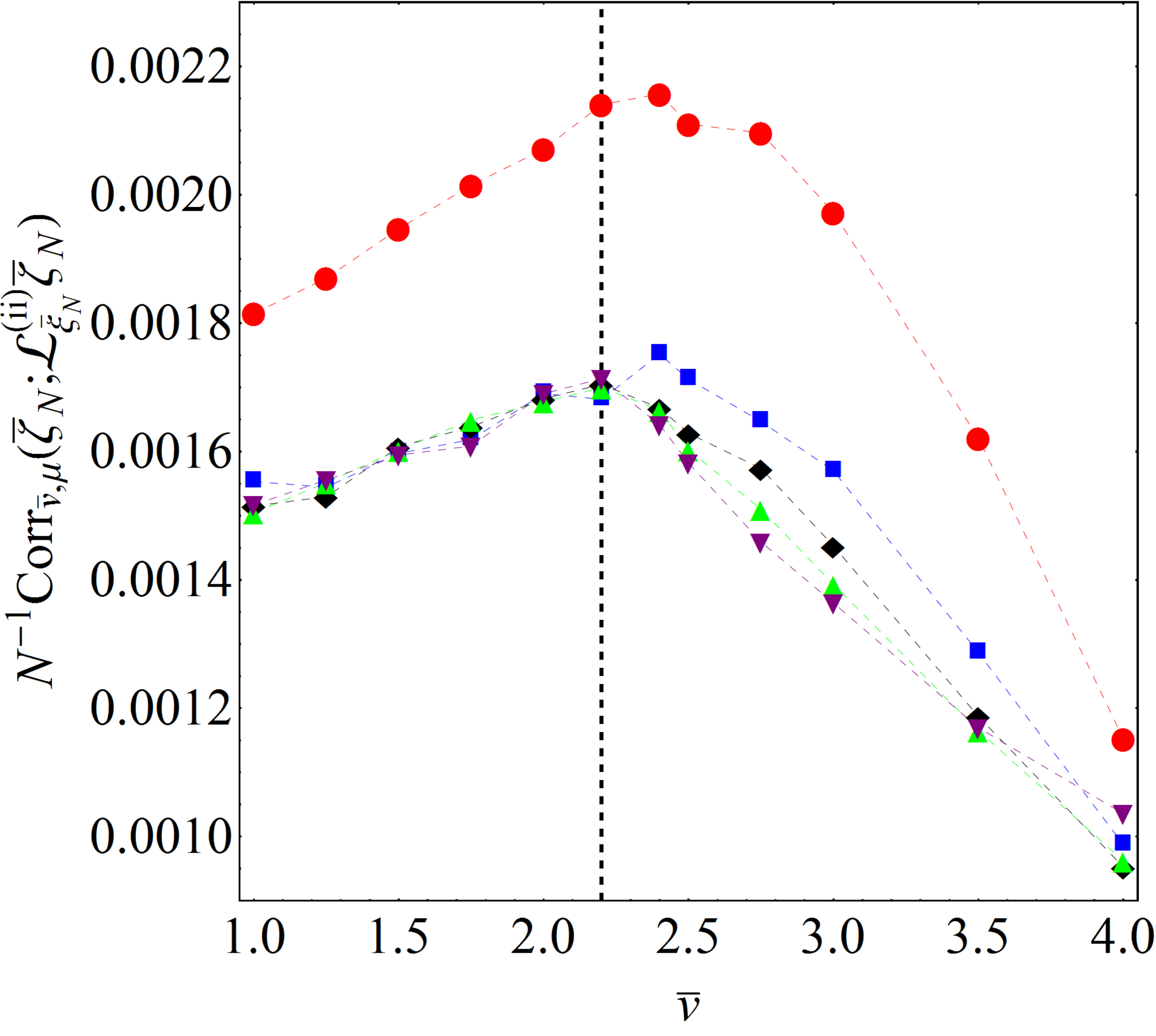}\hskip 1truecm
  \includegraphics[scale=0.2,keepaspectratio=true]{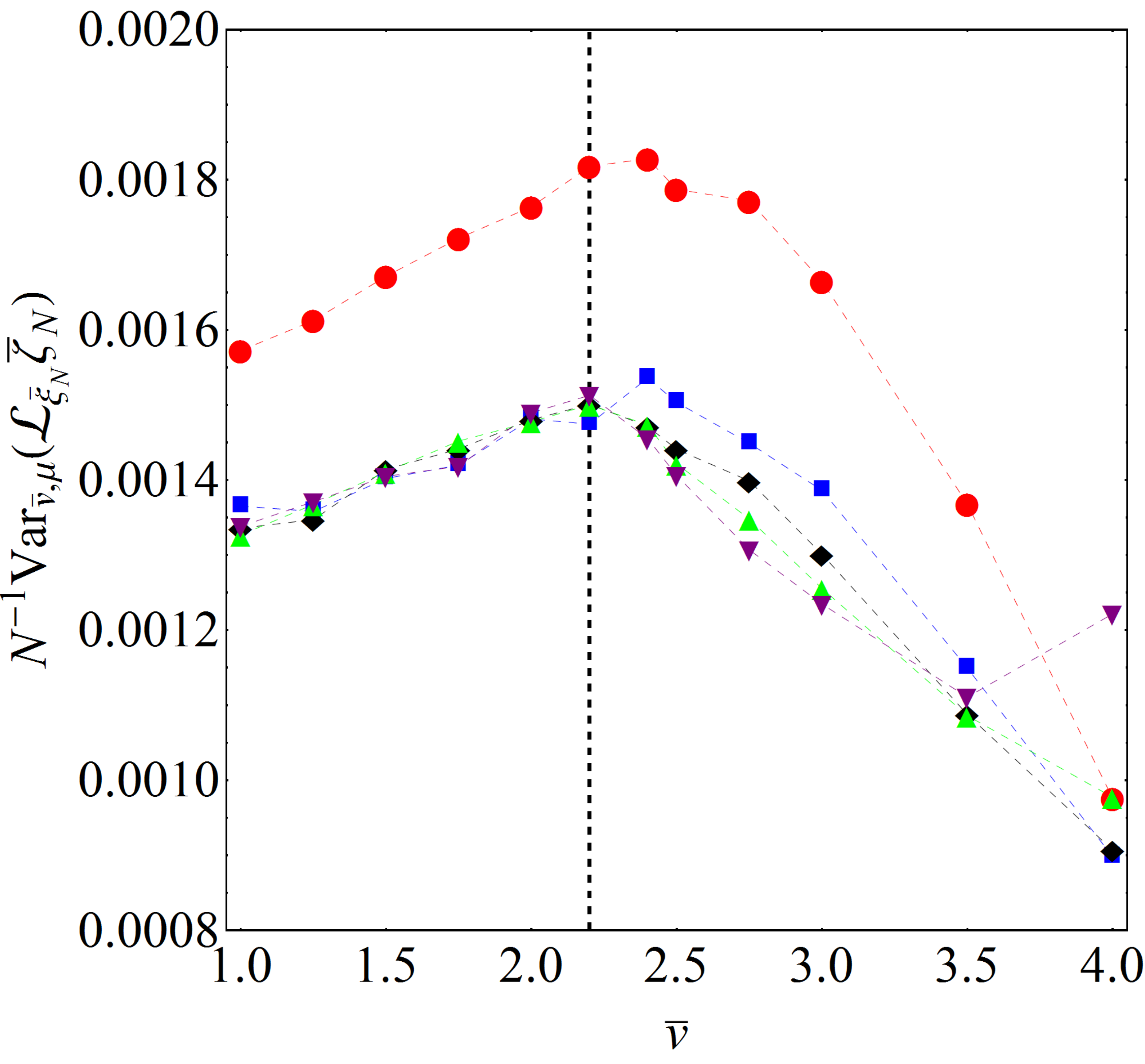}
  \includegraphics[scale=0.2,keepaspectratio=true]{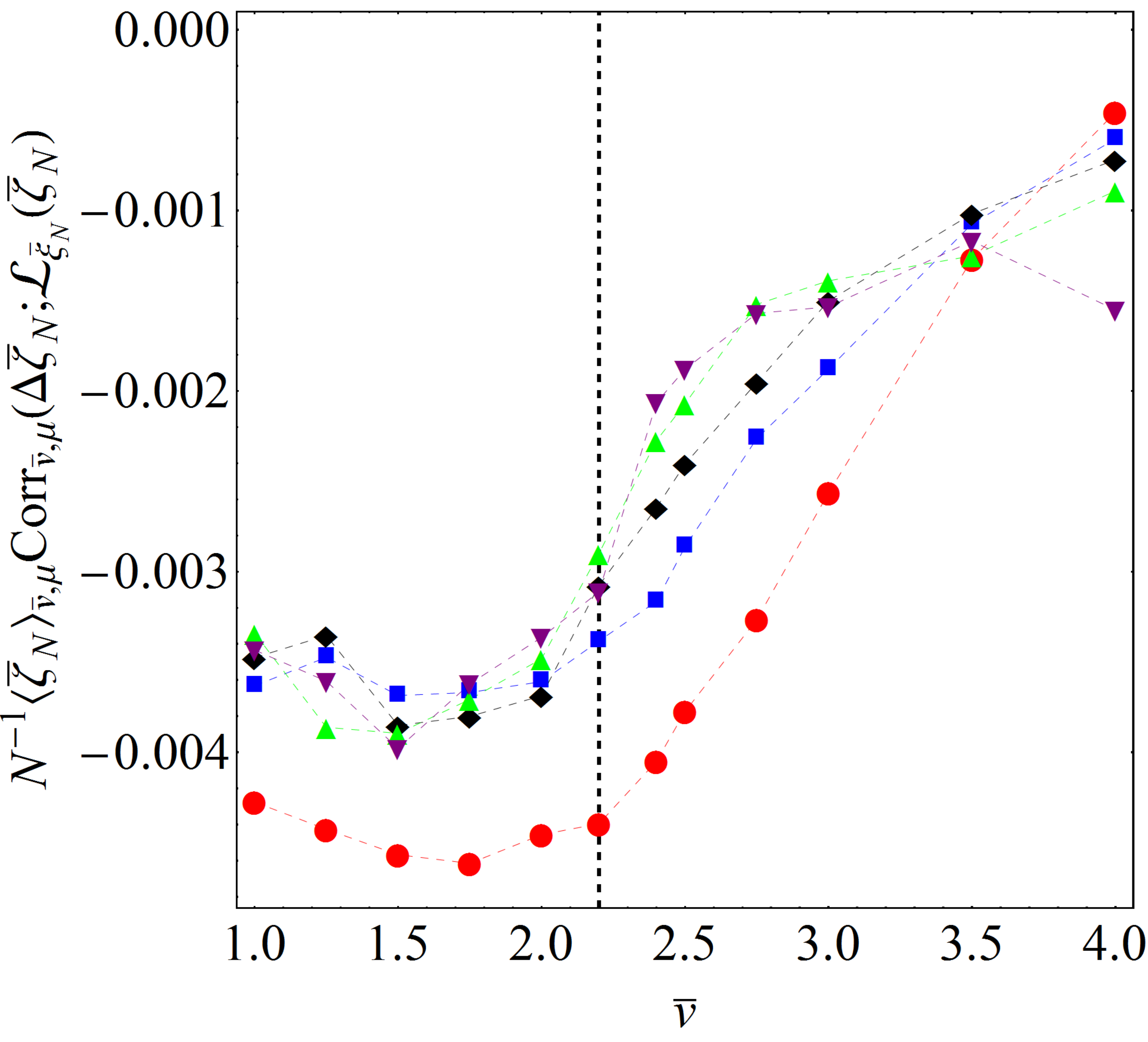}\hskip 1truecm
\includegraphics[scale=0.2,keepaspectratio=true]{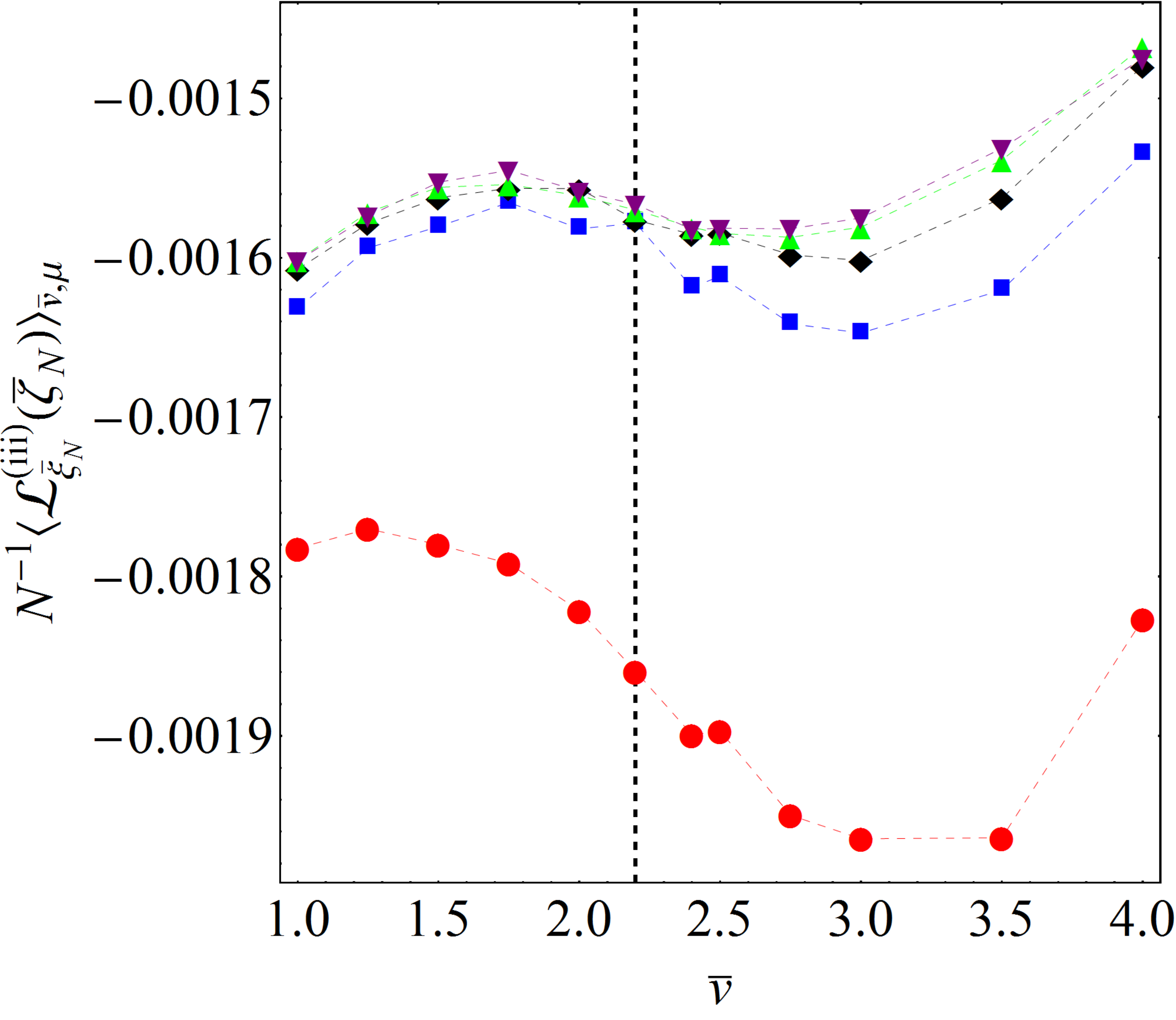}
\caption[10pt]{Quantites entering Eq.\protect\eqref{eq:microcanEntropy_Derivative_Xi} for lattices with different $N$. In particular, $N=100$ (red full circles),$N=400$ (blue squares), $N=900$ (black diamonds), $N=1600$ (green triangles), $N=2500$ (purple reversed triangles). }
\label{LieXi2Zeta}
\end{figure}

\begin{figure}[ht!]
\includegraphics[scale=0.2,keepaspectratio=true]{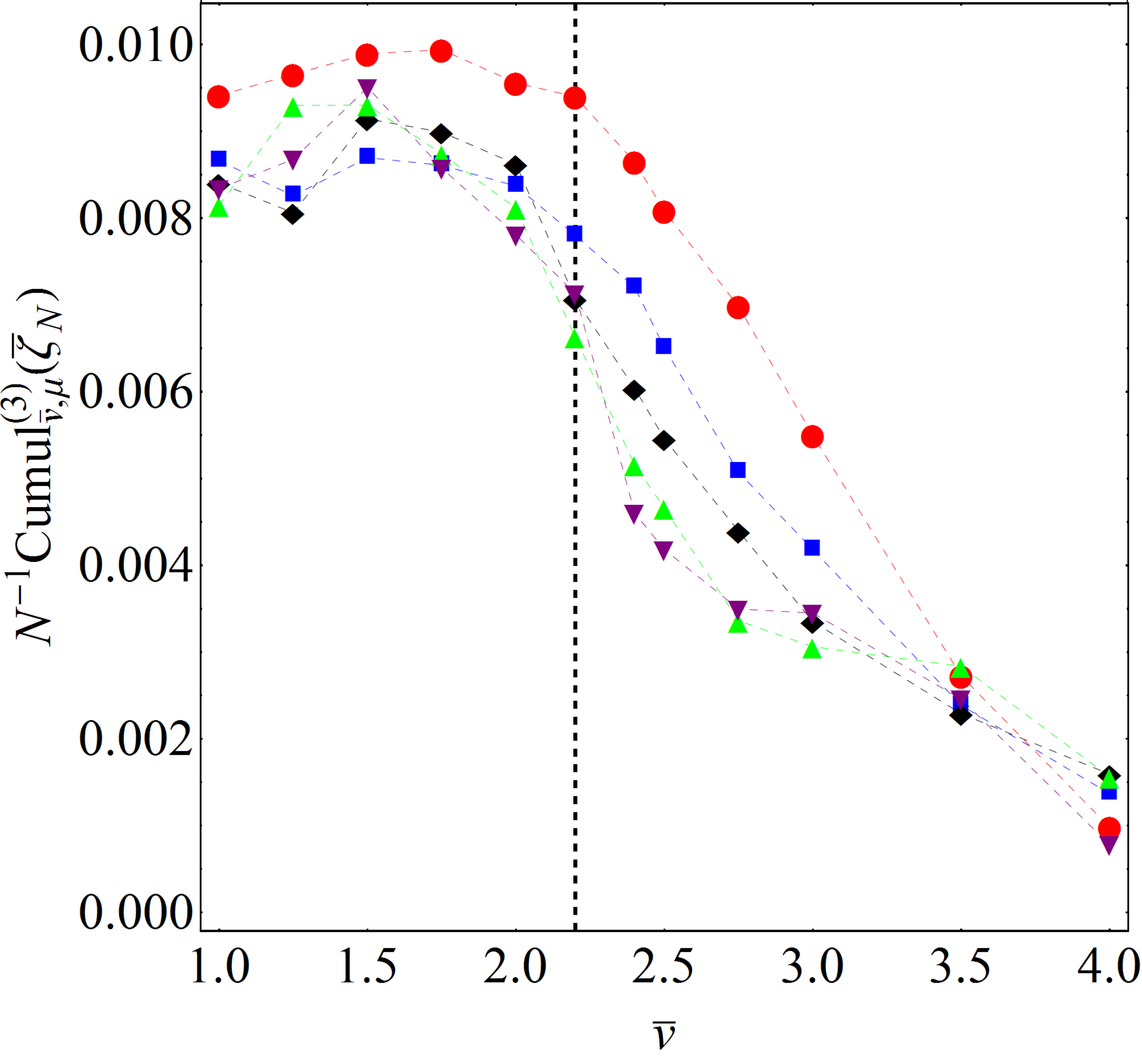}\hskip 1truecm
\includegraphics[scale=0.35,keepaspectratio=true]{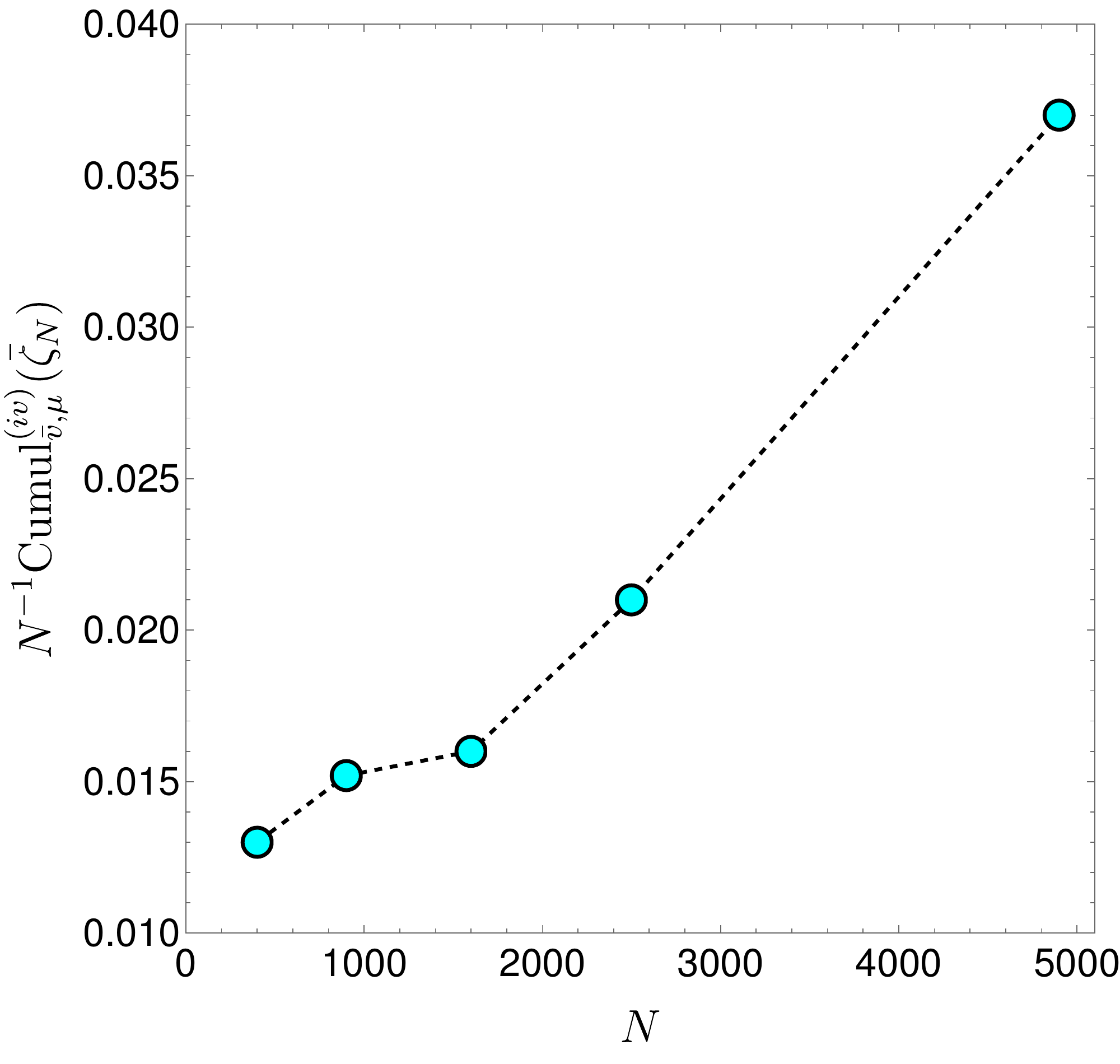}
\caption[10pt]{Quantites entering Eq.\protect\eqref{eq:microcanEntropy_Derivative_Xi} for lattices with different $N$. Left panel: third cumulant of  $\overline{\zeta}_N$ for $N=100$ (red full circles), $N=400$ (blue squares), $N=900$ (black diamonds), $N=1600$ (green triangles), $N=2500$ (purple reversed triangles). Right panel: fourth cumulant of 
$\overline{\zeta}_N$ computed at the transition energy density $\bv_c \simeq 2.2$, here a systematic deviation from the gaussian scaling with $N$ is well evident.}
\label{fgr:D3S_Cumul3Zeta}
\label{Fourcumul}
\end{figure}

The results of numerical simulations are reported for each single single term entering Eq.\eqref{eq:microcanEntropy_Derivative_Xi}. When properly rescaled with $N$, under the hypothesis of  diffeomorphism (at any $N$ and also asymptotically)  of the equipotential hypersurfaces,  all these terms are expected to be uniformly bounded with $N$. Very interestingly, it is found that across  the vertical dashed line denoting the phase transition point at the potential energy density $\bv_c \simeq 2.2$ all these terms do not show any tendency to change with $N$, except for the case $N=10\times 10$ (for which $36\%$ of the total number of sites belong to the boundary, making the finite size effects more relevant). There is only one very significative exception, the fourth cumulant of $\overline{\zeta}_N$ which, computed around the transition value, is found to systematically grow with $N$. This has been computed by means of the relation
\begin{equation}\label{cml4}
\mathrm{Cuml}^{(4)}_{N\bv,\mu}\overline{\zeta}= \frac{d}{dv}\left[ \mathrm{Cuml}^{(3)}_{N\bv,\mu}\overline{\zeta}_N\right]  - 3\left\langle\overline{\zeta}_N\right\rangle_{\overline{v},\mu}\left(\mathrm{Corr}_{\overline{v},\mu}\left(\Delta\overline{\zeta}_N;\mathcal{L}_{\overline{\boldsymbol{\boldsymbol{\xi}}}_N}(\overline{\zeta}_N)\right)\right)
\end{equation}
where the derivative of the third cumulant is evaluated numerically. 

This means that $\overline{\zeta}_N$ is not a Gaussian random process along a MCMC the invariant measure of which is the microcanonical measure.
As $\overline{\zeta}_N$ is proved to be a Gaussian random process under the constraining hypothesis of asymptotic diffeomorphicity of the  level sets 
$\{\Sigma_{N\vb}^{V_N}\}_{\vb\in [\vb_0, \vb_1]}$, the growth with $N$ of  $N^{-1}\mathrm{Cuml}^{(4)}_{N\bv,\mu}\overline{\zeta}$ entails the loss of asymptotic diffeomorphicity among the $\{\Sigma_{N\vb}^{V_N}\}_{\vb<\vb_c}$ and the $\{\Sigma_{N\vb}^{V_N}\}_{\vb>\vb_c}$ for some $\vb_c$. This means that the 2D lattice $\phi^4$ model does not fulfil a basic requirement of the Main theorem formulated in the present work. Therefore, the 2D lattice $\phi^4$ model is not a counterexample to the present version of the topological necessity theorem.

As it has been pointed out in Remark \ref{diffclassS}, a second order phase transition in the microcanonical ensemble is associated with an asymptotic discontinuity of the third derivative of the entropy and hence an asymptotic divergence of the fourth derivative of the entropy
\bigskip\bigskip


\section{Historical overview and challenges of the topological theory of phase transitions}\label{appC}
After the investigation of specific, exactly solvable models \cite{fernando,exact1,exact2,exact3,exact4} corroborating the Topological Hypothesis, during the last decade and a half other systems have been studied, shedding doubts on the general validity of the new theory. Even if all these studies are important to better define the validity limits of the Topological Hypothesis, some authors have too quickly drawn fatal verdicts against it. 

In fact, for the sake of clarity, let us begin by remarking that the proposed theory applies a-priori to systems described by smooth, finite-range potentials such that the level sets  $\Sigma_v^{V_N}$ and their associated balls $M_{v}^{V_N}$ qualify as \textit{good differentiable and compact manifolds}. Moreover, as we have preliminarily discussed in Ref.\cite{vaff}, and as it is thoroughly tackled throughout the present paper, topology changes - as loss of diffeomorphicity among differentiable manifolds - can take place also in the absence of critical points of the potential function.

Thus, coming to specific examples, in \cite{kastner-cazz} the author tackled  a system modelling in one-dimension a localization-delocalization transition of interfaces. 
This so-called Burkhardt model was  also considered 
in \cite{romani1d} where the pinning potential was considered in a modified version. The absence of coincidence between topological and thermodynamical transitions was reported in both Refs. \cite{kastner-cazz} and \cite{romani1d}.

However, the Burkhardt model has two bad properties with respect to the conditions of the Topological Hypothesis, formalized by the theorems in Refs.\cite{book,prl1,NPB1,NPB2} and
by the theorem proved in the present paper. In fact, the pinning potentials considered in
Refs. \cite{kastner-cazz} and \cite{romani1d} are singular as they need infinitely steep potential barriers to constrain the coordinates of the system on a semi-infinite positive line, and the configuration-space submanifolds are noncompact. 

In \cite{grinza}, and also in \cite{romani1d}, another  one dimensional model was considered, the so-called Peyrard--Bishop model describing  DNA denaturation. 
In Ref. \cite{romani1d} the authors considered also a modified version of the Peyrard--Bishop model, and for both versions of the model the topological transition is found at a critical value of the potential energy which does not correspond to critical energy value of the thermodynamic transition.
The configuration space submanifolds corresponding to these models are {\it noncompact}, and the critical manifolds - containing only one critical point of infinite coordinates - are infinitely large. Again, these models are outside the domain of validity of the theorems in Refs.\cite{book,prl1,NPB1,NPB2} and in the present paper.

Another model, allegedly disproving the topological theory,  is the mean-field
Berlin--Kac spherical model. In Ref. \cite{stariolo}  the two cases of zero and non-zero external field were considered. In the former case there is a continuous phase transition and in the latter case there is no phase transition. The two cases did not display much difference when considered from the topological viewpoint.
However, for this model there is a strong statistical ensemble inequivalence, in fact the continuous phase transition for zero external field is only predicted in the framework of canonical ensemble, whereas it is absent in the framework of microcanonical ensemble \cite{kastnerRT,kastner-ineq} which is the reference framework where the topological theory is formulated. Thus there is no contradiction. Moreover, considering that the ergodic invariant measure of Hamiltonian flows is the microcanonical measure, and considering that the objective reality of any physical system is dynamics, the microcanonical ensemble  has
to be considered the fundamental statistical ensemble.

By the way, this is not the only case of this kind of ensemble inequivalence, for example, though working in the opposite way, 
the clustering phase transition in a self-gravitating $N$-body system found in the microcanonical ensemble framework is
completely absent in the canonical ensemble \cite{mnras}.

In \cite{stariolo1} the authors tackle a modified version of the Berlin-Kac model which is constrained with the introduction of a long-range 
correlation among the degrees of freedom.
In spite of the claim that this model is the first case of short-range, confining potential where the phase transition is not entailed by a topology 
change in phase space or configuration space, the spherical constraint, by limiting the freedom of mutual independent variation of all the degrees 
of freedom, makes this model a {\it long-range} interaction one. And even though the spherical constraint is, so to speak, a weak constraint,
It plays a crucial role because without it the model is trivial. 


Another system apparently going against the topological theory is  the mean-field $\phi^4$ model.  The phase transition point of this model does not correspond to the presence of critical points of the potential, that is, it has no topological counterpart \cite{romani,schilling,baroni}.
Some correspondence between the topological and thermodynamic transitions was recovered for this model in Ref. \cite{romani} by introducing a suitable weakening of the Topological Hypothesis. However, the mean-field $\phi^4$ model undergoes a ${\mathbb Z}_2$-symmetry-breaking phase transition,
as in the case of the short range $\phi^4$ model \cite{vaff}. As a consequence, the reason why this system  is not a counterexample of the topological
theory is twofold: on the one side  it violates the condition of asymptotic diffeomorphicity of the level sets  $\Sigma_v^{V_N}$ put forward in the present work, and, on the other side,  in the broken symmetry phase even in the absence of critical points of the potential in the $N\to\infty$ limit the splitting of the configuration space into two disjoint submanifolds is actually a major change of topology in correspondence with the phase transition point. 

All the attempts at falsifying the Topological Hypothesis are useful to better outline the domain of validity of the theory, which, on the other hand, is not intended to apply to {\it all} the possible phase transitions. The theory can leave outside its validity domain models with unbound and long-range potentials  without being invalidated by this kind of systems.

Summarizing, the Topological Hypothesis is coherent and now free of counterexamples, nevertheless the above 
mentioned alleged counterexamples have the merit of showing that some - perhaps much - work remains to be done, mainly in the case of 
long-range interactions.
In fact, for example, for the exactly solvable model in Refs.\cite{exact2,exact3}, which is a  mean-field XY model, thus with long-range
interactions, a sharp topological transition in configuration space is clearly at the origin of the phase transition. At variance with the
mean-field $\phi^4$ model described by polynomial potentials, the mean-field XY model is described by a potential bounded
from above. Whether and why this fact could explain the different conformity of these models to the topological description of
their transitional behaviour is still a wide open question.

\section{Concluding remarks} 
The present work is a substantial leap forward of the topological theory of phase transitions which was seriously undermined by the counterexample mentioned in the previous sections. The theory is rooted in the study of thermodynamical phase transitions from the viewpoint of microscopic Hamiltonian dynamics. As Hamiltonian flows can be identified with geodesics flows of suitable differentiable manifolds, it turned out that across a phase transition point these manifolds undergo major geometrical changes of topological origin. This is to remark that topology is, so to speak, naturally implied by the fundamental/dynamical approach and it is not just conjectured to play a role.  
The first important consequence of this approach is that the occurrence of a phase transition is not the consequence of a loss of analyticity of statistical measures but it is already encoded in the potential function describing the interactions among the degrees of freedom of a system. This makes the \textit{thermodynamic limit dogma} no longer necessary neither from the conceptual side nor from the mathematical description side. And, of course, this is interesting when tackling phase transition phenomena in mesoscopic and nanoscopic systems. Moreover, phase transitions phenomena in the absence of symmetry-breaking - and thus in the absence of an order parameter - have been successfully tackled in the topological framework, at present, at least  in the case of a model with a gauge symmetry \cite{dualising}, for a 2D-model with an $O(2)$ symmetry undergoing a Kosterlitz-Thouless transition \cite{kosterlitz}, and for the protein folding transition \cite{proteinfold}.
It is worth mentioning that in a recent paper \cite{loris} - partly based on some results given in \cite{gori2022} - a purely geometric theory of phase transitions has been put forward. In this work it is proposed that 
Bachmann’s classification of phase transitions \cite{bachmann,PRL-Bachmann} for finite-size systems can be reformulated in terms of geometric
properties of the energy level sets associated to a given Hamiltonian function; here the energy-derivatives of the entropy are associated to specific combinations of geometric curvature properties of the energy level sets. There is no contradiction between the geometric theory and the topological theory
of phase transitions, mainly because sharp changes of the geometry of the leaves (energy level sets) of a foliation of phase space can be generically attributed to deeper changes of topological kind. However, the precise relationship between geometry and topology is given by theorems in differential topology and, unfortunately, there is only a few number of these theorems that can be constructively used (essentially the Gauss-Bonnet-Hopf, the Chern-Lashof, and Pinkall theorems \cite{book}). Therefore the geometric approach has some practical advantage with the respect to the topological one in what curvature properties of the energy level sets can be always explicitly computed.

It is noteworthy that, in principle, the topological approach to classical phase transitions - addressed in the present work,  as well as the geometric approach in Ref.\cite{loris} - can be extended to the treatment of quantum transitions by means of  Wick's analytic prolongation to imaginary times of the  path-integral generating functional of quantum field theory, this allows to map a quantum system onto a formally classical one described by a classical partition function written with the euclidean Lagrangian action, on lattice to have a countable number of degrees of freedom \cite{book}. 

Finally, recent developments of powerful computational methods in algebraic topology, like those of \textit{persistent homology} \cite{Carlsson,noi}, provide the topological description of phase transitions with new useful constructive tools in addition to the existing concepts and methods of differential topology.

\bigskip\bigskip
\subsection*{Appendix A. Uniform boundedness of ${\mathscr R}$}\label{appA}
Let us now show how asymptotic diffeomorphicity entails uniform boundedness with $N$ of the Ricci scalar curvature defined in equation \eqref{scalar1}, and using $\Vert\boldsymbol{\xi}\Vert = \Vert\nabla V(\textit{\textbf{q}})\Vert^{-1} $, 
\begin{equation} 
{\mathscr R} = \frac{1}{N(N-1)}\left\{ -\triangle\log  \Vert\boldsymbol{\xi}\Vert^{-1}+ \nabla\cdot \left[ \triangle V(\textit{\textbf{q}})\  \boldsymbol{\xi}   \right] \right\} \ ,    
\label{scalar1-append}
\end{equation} 
where the second term in the r.h.s. is 
\begin{eqnarray}
\nabla\cdot \left[ \triangle V(\textit{\textbf{q}})\  \boldsymbol{\xi}   \right] &=& [ \nabla\triangle V(\textit{\textbf{q}})]\cdot  \boldsymbol{\xi}  + [\triangle V(\textit{\textbf{q}})] \nabla\cdot\boldsymbol{\xi}\label{array1a}\\
&=& \triangle V(\textit{\textbf{q}})\  \partial^i\left(\frac{\partial_i V(\textit{\textbf{q}})}{\Vert\nabla V(\textit{\textbf{q})}\Vert^2}\right) + \frac{\partial_j V(\textit{\textbf{q})}}{\Vert\nabla V(\textit{\textbf{q})}\Vert^2}\partial^j\partial^k\partial_kV(\textit{\textbf{q}})\label{array1b}
\end{eqnarray}
and using \cite{convenzione}
\begin{eqnarray}
 \triangle V(\textit{\textbf{q}}) &=& \nabla\cdot[\nabla V(\textit{\textbf{q}})] = \nabla\cdot[\Vert\nabla V(\textit{\textbf{q}})\Vert^2  \boldsymbol{\xi} ] = \partial^i \left(\frac{\xi_i}{\Vert\boldsymbol{\xi}\Vert^2 }\right)\\
 &=&\frac{1}{\Vert\boldsymbol{\xi}\Vert^2 } (\partial^i\xi_i)  - \frac{4}{\Vert\boldsymbol{\xi}\Vert^2}  \frac{\xi_i}{\Vert\boldsymbol{\xi}\Vert }\frac{\xi^j}{\Vert\boldsymbol{\xi}\Vert } (\partial^i\xi_j)     \label{array2b}  
\end{eqnarray}
after Eqs.\eqref{normaxi},\eqref{norma}, \eqref{eq:AsympDiffCk} all these terms are uniformly bounded in $N$, and so does $\nabla\cdot\boldsymbol{\xi}$, moreover, the denominator 
$\Vert\boldsymbol{\xi}\Vert^{-4}\sim N^2$ is compensated by the pre-factor $1/N(N-1)$. 
therefore the second term in the r.h.s. of Eq.\eqref{array1a} is also uniformly bounded in $N$. Then the first term of Eq.\eqref{array1a} is obtained by applying the operator $\boldsymbol{\xi}\cdot\nabla$ to Eq.\eqref{array2b}, and, after trivial algebra of the same kind of that leading to Eq.\eqref{array2b}, one obtains a lengthy expression - containing mixed second order derivatives of the components of $\boldsymbol{\xi}$ - which are uniformly bounded under the assumption of asymptotic diffeomorphicity. On the other hand, for smooth and regularized potentials, if $n$ is the coordination number of the potential, and ${\textgoth  m}_0$ is the maximum value of $\partial^i\partial_i V$, then $ \triangle V(\textit{\textbf{q}})$ is bounded by 
$n\, {\textgoth m}_0N/[N(N-1)]$. By the same token, if ${\textgoth  m}_1$ is the maximum value of $\partial_j\partial^i\partial_i V$ then $\boldsymbol{\xi}\cdot\nabla\triangle V(\textit{\textbf{q}})$ is uniformly bounded by $n\, {\textgoth m}_1 B$, where $B$ is the constant of Eq.\eqref{eq:AsympDiffCk}.

Now, coming to the first term in the r.h.s. of Eq.\eqref{scalar1-append}, that is $\partial^i\left( \partial_i \log \Vert \nabla V(\textit{\textbf{q}}) \Vert\right)$, we have
\begin{eqnarray}
\partial^i\left( \partial_i \log  \Vert\boldsymbol{\xi}\Vert^{-1}\right) &= & \partial^i(\Vert \boldsymbol{\xi}\Vert \partial_i\Vert\boldsymbol{\xi}\Vert^{-1}) = \partial^i\Vert \boldsymbol{\xi}\Vert \partial_i\Vert\boldsymbol{\xi}\Vert^{-1} + \Vert \boldsymbol{\xi}\Vert \  \partial^i\partial_i\Vert\boldsymbol{\xi}\Vert^{-1}\\
&=& - \frac{1}{\Vert\boldsymbol{\xi}\Vert^2 } (\partial_i \Vert \boldsymbol{\xi}\Vert )^2 + \left[ - \frac{1}{\Vert\boldsymbol{\xi}\Vert} (\partial^i \partial_i \Vert \boldsymbol{\xi}\Vert )  + \frac{2}{\Vert\boldsymbol{\xi}\Vert^2 }  (\partial_i \Vert \boldsymbol{\xi}\Vert )^2\right]\\
&=& - \frac{1}{\Vert\boldsymbol{\xi}\Vert} (\partial^i \partial_i \Vert \boldsymbol{\xi}\Vert ) + \frac{1}{\Vert\boldsymbol{\xi}\Vert^2 } (\partial_i \Vert \boldsymbol{\xi}\Vert )^2\label{a8}
\end{eqnarray}
and
\begin{equation} 
\frac{1}{\Vert\boldsymbol{\xi}\Vert^2 } (\partial_i \Vert \boldsymbol{\xi}\Vert )^2 = \frac{1}{\Vert\boldsymbol{\xi}\Vert^2 } (\partial_i \sqrt {\xi_j\xi^j})^2 = \frac{\xi_j\xi^j}{\Vert\boldsymbol{\xi}\Vert^4 } (\partial_i\xi_j)^2
\end{equation}
which is uniformly bounded after Eqs.\eqref{normaxi},\eqref{norma},\eqref{eq:AsympDiffCk} and the pre-factor $1/N(N-1)$. Then for the first term in Eq.\eqref{a8}  we get
\begin{equation} 
 \frac{1}{\Vert\boldsymbol{\xi}\Vert} (\partial^i \partial_i \Vert \boldsymbol{\xi}\Vert ) = \frac{1}{\Vert\boldsymbol{\xi}\Vert} \partial^i \left( \frac{\xi^j}{\Vert\boldsymbol{\xi}\Vert}\, \partial_i\xi_j\right) = 
\frac{1}{\Vert\boldsymbol{\xi}\Vert}\left[  \frac{\xi^j}{\Vert\boldsymbol{\xi}\Vert}\, \partial^i\partial_i\xi_j +  \frac{(\partial_i\xi_j)^2}{\Vert\boldsymbol{\xi}\Vert} -  \frac{\xi_j\xi^j\, (\partial_i\xi_j)^2}{\Vert\boldsymbol{\xi}\Vert^3}  \right]
\end{equation}
which, under the same conditions mentioned above, is also uniformly bounded in $N$. The relevant consequence is that under the assumption of asymptotic diffeomorphicity of potential energy level sets, the scalar Ricci curvature ${\mathscr R}$ in Eq.\eqref{scalar1-append} is uniformly bounded in $N$ and so do all the principal curvatures of the manifolds transformed under the action of the vector field $\boldsymbol{\xi}$.

\subsection*{Appendix B. Lie derivatives of the vector field $\boldsymbol{\xi}$}\label{appB}
In the following  we derive explicit expressions of the Lie derivatives of the one-parameter diffeomorphism-generating vector field $\boldsymbol{\xi}$ for a potential $V$ in "critical points-free" regions of configuration space $(\mathcal{X},g_{\mathbb{R}^N}$ endowed with a Riemannian metric. Let $(q_1,....,q_N)$ be a set of coordinates in configuration space. In what follows we shall refer to  $\partial_i=\partial/\partial q^i$ so that
$(\nabla V)_i =\partial_i V$ and the Hessian $(\Hess V)_{ij}=\partial^2_{ij}V$.\\
With these chioces the divergence of the vector field $\zeta=\mathrm{div}_{\mathbb{R}^N}\boldsymbol{\xi}$ reads:
\begin{equation}
\mathrm{div}_{\mathbb{R}^N}\boldsymbol{\xi}=\dfrac{\Delta V}{\|\nabla V\|^2}-2\dfrac{\nabla V \cdot(\Hess V\nabla V)}{\|\nabla V\|^4}
\end{equation}
where $\Delta(\cdot)=\sum_{i}^N \partial^{i}\partial_{i}(\cdot)$ is the Laplacian operator in the Euclidean configuration space and $\|\boldsymbol X\|^2=g_{\mathbb{R}^N}(\boldsymbol X,\boldsymbol X)$ is the Euclidean norm.  As the Lie derivative operator along the flow generated by the vector field $\boldsymbol{\xi}$ is 
\begin{equation}
\mathcal{L}_{\boldsymbol{\xi}}(\cdot)=(\boldsymbol{\xi}\cdot\nabla)(\cdot)=\sum_{i=1}^{N}\dfrac{\partial^{i}V}{\|\nabla V\|^2}\partial_i(\cdot)
\end{equation}
the first derivative reads
\begin{equation}\label{derivLie}
\begin{split}
\mathcal{L}_{\boldsymbol{\xi}}(\zeta)=&\dfrac{\nabla V\cdot\nabla(\Delta V)}{\|\nabla V\|^4}-2\dfrac{(\nabla V \cdot\Hess(V)\nabla V)\Delta V+2\|\Hess V\nabla V\|^2+\mathrm{D}^3V(\nabla V,\nabla V,\nabla V)}{\|\nabla V\|^6}+\\
&+8\dfrac{(\nabla V \cdot\Hess V\nabla V)^2}{\|\nabla V\|^8}
\end{split}
\end{equation}
where $\mathrm{D}^3V(\nabla V,\nabla V,\nabla V) = \de^3_{ijk}V \de_i V \de_jV \de_k V $.

The second order derivative (with the aid of symbolic manipulation with Mathematica) reads
\begin{eqnarray}\label{deriv2Lie}
&&\mathcal{L}_{\boldsymbol{\xi}}^{(ii)}(\zeta)= \dfrac{\nabla(\Delta V)\cdot (\Hess V \nabla V)+\nabla V \cdot(\Hess(\Delta V)\nabla V)}{\|\nabla V\|^6}+\nonumber\\
&-&2\Biggr[\dfrac{\Delta V\mathrm{D}^3 V(\nabla V,\nabla V,\nabla V)+2\Delta V\|\Hess V\nabla V\|^2+4(\Hess V\nabla V)\cdot(\Hess V\Hess V\nabla V)}{\|\nabla V\|^8}+\nonumber\\
&+&\dfrac{7\mathrm{D}^3V(\Hess V\nabla V,\nabla V,\nabla V)+\mathrm{D}^4V(\nabla V,\nabla V,\nabla V,\nabla V)+3 (\nabla V\cdot \Hess V\nabla V)(\nabla V\cdot \nabla(\Delta V))}{\|\nabla V\|^8}\Biggr]+\nonumber\\
&+&\dfrac{28(\nabla V\Hess V\nabla V)\left[2\|\Hess V \nabla V\|^2+\mathrm{D}^3V(\nabla V,\nabla V,\nabla V)\right]+12(\nabla V\Hess V\nabla V)^2\Delta V}{\|\nabla V\|^{10}}+\nonumber\\
&-&64\dfrac{(\nabla V\Hess V\nabla V)^3}{\|\nabla V\|^{12}}
\end{eqnarray}
\vfill
and the third order derivative (with the aid of symbolic manipulation with Mathematica) reads

\begin{equation}\label{deriv3Lie}
\begin{split}
&\mathcal{L}_{\boldsymbol{\xi}}^{(iii)}(\zeta)=\dfrac{3 \nabla V \cdot\Hess(\Delta V)\Hess V\nabla V+\mathrm{D}^3\Delta V(\nabla V,\nabla V,\nabla V)+\mathrm{D}^{3}V(\nabla V,\nabla V,\nabla(\Delta V))}{\|\nabla V\|^8}+\\
&+\dfrac{\nabla (\Delta V)\cdot \Hess V\Hess V \nabla V}{\|\nabla V\|^8}-2\Biggr[\dfrac{4 \mathrm{D}^{3}V(\nabla V,\nabla V,\nabla V)(\nabla V \cdot \nabla(\Delta V))}{\|\nabla V\|^{10}}+\\
&+\dfrac{
7 \mathrm{D}^{4}V(\nabla V,\nabla V,\nabla V,\Hess V\nabla V)}{\|\nabla V\|^{10}}+\\
&+\dfrac{15 \mathrm{D}^{3}V(\nabla V,\nabla V,\Hess V \Hess V \nabla V)+7\|\mathrm{D}^{3}V(\nabla V,\nabla V)\|^2+18 \mathrm{D}^{3}V(\Hess V\nabla V,\Hess V \nabla V,\nabla V)}{\|\nabla V\|^{10}}+\\
&+\dfrac{4\mathrm{D}V(\nabla V,\nabla V, \nabla V,\Hess V \nabla V)+\mathrm{D}^{5}V(\nabla V, \nabla V, \nabla V,\nabla V, \nabla V)+8(\nabla V \cdot \nabla (\Delta V))\|\Hess V \nabla V\|^2}{\|\nabla V\|^{10}}+\\
&+\dfrac{8\|\Hess V \Hess V\nabla V\|^2+7\mathrm{D}^3 V(\nabla V,\nabla V,\Hess V \nabla V)\Delta V}{\|\nabla V\|^{10}}+\\
&\dfrac{\Delta V\mathrm{D}^{4}V(\nabla V,\nabla V,\nabla V,\nabla V)+4\Delta V (\Hess V \nabla V)\cdot \Hess V \Hess V \nabla V}{\|\nabla V\|^{10}}+\\
&+\dfrac{6(\Hess V \cdot \Hess V \nabla V)(\nabla V\cdot \Hess (\Delta V)\nabla V)+6(\Hess V \cdot \Hess V \nabla V)(\nabla(\Delta V)\cdot \Hess V \nabla V)}{\|\nabla V\|^{10}}\Biggr]+\\
&+4\Biggr[\dfrac{7\left(\mathrm{D}^3 V(\nabla V,\nabla V, \nabla V)\right)^2+28 \mathrm{D}^{3}V(\nabla V, \nabla V, \nabla V) \|\Hess V\nabla V\|^2}{\|\nabla V\|^{12}}+\\
&+\dfrac{10 \Delta V \mathrm{D}^3 V(\nabla V, \nabla V,\nabla V)(\nabla V \cdot \Hess V \nabla V)}{\|\nabla V\|^{12}}+\\
&+\dfrac{28 \|\Hess V \nabla V\|^4+20 \Delta V\|\Hess V \nabla V\|^2(\nabla V\cdot \Hess V \nabla V)}{\|\nabla V\|^{12}}+\\
&+\dfrac{(\nabla V\cdot \Hess V \nabla V)[77 \mathrm{D}^3 V(\nabla V,\nabla V,\Hess V \nabla V)}{\|\nabla V\|^{12}}+\\
&+\dfrac{11 \mathrm{D}^4 V(\nabla V,\nabla V,\nabla V,\nabla V)+44 (\Hess V \nabla V)\cdot(\Hess V \Hess V \nabla V)}{\|\nabla V\|^{12}}+\\
&+\dfrac{15 (\Hess V \cdot \Hess V \nabla V)(\nabla V\cdot \nabla(\Delta V))]}{\|\nabla V\|^{12}}\Biggr]+\\
&-8\Biggr[\dfrac{59\mathrm{D}^{3}V(\nabla V,\nabla V,\nabla V)(\nabla V \cdot \Hess V \nabla V)^2}{\|\nabla V\|^{14}}+\\
&+\dfrac{(\nabla V  \cdot \Hess V \nabla V)^2 [118 \|\Hess V \nabla V\|^2+15 \Delta V (\nabla V  \cdot \Hess V \nabla V)]}{\|\nabla V\|^{14}}\Biggr]+768\dfrac{(\nabla V  \cdot \Hess V \nabla V)^4}{\|\nabla V\|^{16}}
\end{split}
\end{equation}

\vfil

\section{Acknowledgments}
This work has been done within the framework of the project MOLINT which has received funding from the Excellence Initiative of Aix-Marseille University - A*Midex, a French ``Investissements d'Avenir'' programme.
This work was also partially supported by the European Union’s Horizon 2020 Research and Innovation Programme under grant agreement no. 964203 (FET-Open LINkS project).
Roberto Franzosi acknowledges support by the QuantERA ERA-NET Co-fund 731473 (Project Q-CLOCKS) and the support by the National Group of Mathematical Physics (GNFM-INdAM).
Matteo Gori thanks the financial support of DARPA (USA) for his long term visit at Howard University at Washington D.C. during which part of this work was done.



\begin{thebibliography}{99}
\bibitem{clt-weak} Consider a family of random variables $\{ X_k\}$, if $X_i$ and $X_j$ satisfy finite-range dependence, i.e. there is a positive integer $n_0$ such that if $\vert i - j\vert > n_0$ then $X_i$ and $X_j$ are independent. A law of large numbers for $X_i$ can then be easily proved.
\end{thebibliography}
\end{document}